\documentclass[
  aip,
  jmp,
 amsmath,amssymb,
 reprint,%
]{revtex4-1}

\usepackage{amsmath}
\usepackage{amssymb}
\usepackage{dsfont}

\usepackage{graphicx}
\usepackage{bm,bbm}
\usepackage{color}
\usepackage{placeins}
\usepackage[normalem]{ulem}
\usepackage{soul}
\DeclareMathOperator{\sech}{sech}

\newcommand{\ie}{\textit{i.e.} }
\newcommand{\rmd}{\mathrm{d}}
\newcommand{\rme}{\mathrm{e}}
\newcommand{\rmi}{\mathrm{i}}

\newcommand{\tr}{{\mathrm{tr}}}
\newcommand{\One}{\openone}
\newcommand{\eps}{\varepsilon}
\newcommand{\h}{\hat}
\newcommand{\rmw}{\mathrm{w}}

\newcommand{\la}{\langle}
\newcommand{\ra}{\rangle}

\newcommand{\bA}{{\bf A}}

\newcommand{\bPh}{\tilde{\Phi}}

\newcommand{\abs}[1]{\left|#1\right|}

\newcommand{\mc}[1]{\mathcal{#1}}

\begin{document}

\author{P. C. L\' opez V\' azquez}
\author{L. M. Piñuelas}
\affiliation{Departamento de Ciencias Naturales y Exactas, Universidad de
Guadalajara,
Carretera Guadalajara - Ameca Km. 45.5 C.P. 46600. Ameca, Jalisco, M\'exico.}

\author{G. Montes}
\affiliation{Departamento de F\'isica, Universidad de Guadalajara,
Boulevard Marcelino Garc\'ia Barragan y Calzada Ol\'impica, C.P. 44840,
Guadalajara, Jalisco, M\'exico}

\title[Probing entanglement of a continuous basis system]
{Probing entanglement of a continuous basis system}

\begin{abstract}
 In this paper, we propose a method to probe entanglement in a theoretically
 inaccessible quantum system with either a discrete or continuous basis. 
 Our approach leverages insights into the entanglement distribution within 
 a four-partite quantum system, comprising two qubit-oscillator subsystems with 
 dephasing interactions between each qubit-oscillator pair. The method involves 
 measurements applied only to the accessible two-qubit subsystem, enabling a
 qualitative detection and characterization of quantum 
 correlations in the inaccessible two-oscillator subsystem. This approach provides 
 a novel framework for probing entanglement in continuous-basis systems
 where traditional measures are often 
 inapplicable due to their complexity. Our findings also reveal an intriguing 
 conservative flow-like behavior in the redistribution of entanglement among 
 subsystems, suggesting that entanglement may exhibit conservative properties 
 in pure composite quantum systems.
\end{abstract}
\maketitle

%\tableofcontents
\section{Introduction}

Quantum entanglement stands as one of the most intriguing phenomena in modern physics, 
challenging classical intuitions while enabling revolutionary applications in quantum computing, 
secure communication, and information processing~\cite{Einstein35,Schrodinger35,Nielsen2010}. 
This non-local correlation between subsystems 
of a composite quantum system lies at the heart of quantum advantage, fueling advancements 
such as quantum teleportation, superdense coding, and error-corrected 
computation~\cite{Bennett92,Bouwmeester97}.

Quantifying entanglement between subsystems remains a central task 
in quantum information science, with well-established measures for discrete-variable systems. 
For two-qubit systems, Wootters' concurrence provides a reliable metric for pure states 
and their extension to mixed states through the use of the convex roof~\cite{Wootters98,Wo01}. 
Extending this to higher-dimensional multipartite systems turns out to be very complex, due to the 
intricate structure of multipartite entanglement. This has resulted in a wide variety of 
measures capable of capturing different aspects of multipartite entanglement~\cite{CoKuWo00, PlVi07}. 

For bipartite systems $|\Psi_{\text{AB}}\rangle$, 
the I-concurrence based on the universal inversion superoperator~\cite{RuBuCaHiMi01}, 
generalizes the well-established Wootters’ concurrence for two qubits, however for bipartite 
systems of arbitrary dimensions, the I-concurrence connects the degree of mixture of the 
sub-partitions to the degree of entanglement among them.

Other approaches include geometric measures~\cite{Wei03}, which quantify entanglement by the minimal 
distance to separable states, and genuine multipartite entanglement (GME) 
criteria~\cite{Seevinck08,Guhne09}, such as the minimal bipartite concurrence across partitions. 
For three-qubit systems, the tangle~\cite{Coffman00} captures GME and illustrates entanglement 
monogamy~\cite{Osborne06}. While mixed-state extensions via convex roof~\cite{Uhlmann00,Wootters98} 
are often intractable, entanglement witnesses~\cite{Guhne09,Horodecki96} and the negativity 
measure~\cite{Peres96,Horodecki96,Vidal02} offer more practical alternatives, despite computational 
challenges in large systems. Generalizations include the multipartite concurrence~\cite{Mintert04,Mintert05}, 
Q-concurrence based on Tsallis entropy~\cite{Kim10}, and extensions of negativity to continuous variables~\cite{Adesso04}.

The challenge escalates in continuous variable (CV) systems 
due to the infinite-dimensional Hilbert space~\cite{Adesso07}. 
However, for Gaussian states, powerful tools 
like the PPT criterion~\cite{Simon00}, logarithmic negativity~\cite{Vidal02,Adesso07}, and 
Gaussian entanglement of formation~\cite{Giedke03} allow 
efficient characterization and quantification. These have been instrumental in protocols 
like CV teleportation~\cite{Braunstein98,Loock00,Furusawa98,Yonezawa2004}
and quantum key distribution~\cite{Grosshans2003,Navascues05,Grosshan05}.

For non-Gaussian states, entanglement is harder to capture, but advances include witness-based 
methods~\cite{Hyllus06,Jayachandran23,Shahandeh14}, 
non-positivity of partial transposition (NPT) tests~\cite{Shchukin06},
and emerging resource theories for non-Gaussianity with relevance to 
quantum computation~\cite{Takagi18}.
Despite these developments, entanglement detection and characterization in multipartite CV 
systems remains an active research frontier. Understanding entanglement distribution, governed 
by principles such as monogamy, is essential for applications in quantum 
networks~\cite{Kimble2008,Pirandola2017,Cuquet09}, where 
entanglement acts as a finite resource which can be 
shared, transferred and transformed among multiple parts of the composite system, 
including interaction with auxiliary systems.

In this work, we introduce an indirect measurement protocol to probe entanglement in 
theoretically inaccessible continuous-variable systems by leveraging their coupling to a 
discrete-variable accessible quantum probe. Our method focuses on a four-partite system comprising two 
qubit-oscillator pairs with a dephasing type of coupling between a single qubit to a 
single  oscillator~\cite{Xiang13,Pirkkalainen15}.

The full integrability of the setup~\cite{Lo18, Lo20, Lo21} permit us 
to analyze its solutions under varied initial conditions; 
we demonstrate that standard qubit measurements alone suffice 
to identify entangled versus separable oscillator states. 

This approach enables both qualitative detection and characterization 
of entanglement in the two-oscillator subsystem, even for non-Gaussian or highly 
complex states, offering a novel framework where standard measures are inapplicable.
To qualitatively assess entanglement, we prepare two copies of the system, in a first 
copy, the two-qubit subsystem is initialized in a Bell state, and its concurrence is monitored,
in the second copy the qubits are decoupled but maximally superposed, and the fidelity amplitudes 
(qubit coherence functions) of each qubit-oscillator pair are tracked.
By comparing concurrence dynamics in a Bell-state-prepared copy with 
fidelity amplitudes (qubit coherences) in a decoupled copy, we establish a 
separability criterion: their exact match implies separable oscillators, 
while deviations reveal entanglement and correlation redistribution

Key to our method is the observed conservative redistribution of entanglement among 
subsystems, suggesting that entanglement in pure composite systems may exhibit 
flow-like behavior akin to a conserved quantity. 
This insight guides our measurement 
strategy to propose a quantitative characterization of entanglement in the two-oscillator 
subsystem; moreover, the conservative-like flux of correlations
opens new questions about entanglement dynamics in hybrid quantum systems. 
Our results hold promise for quantum information applications, particularly in scenarios 
where continuous-variable entanglement is essential but direct measurement is infeasible.

We analyze the composite system dynamics using the chord (characteristic) 
function representation of the two-oscillator subsystem~\cite{Ozo98, Ozo02, Oz04}, 
a phase-space representation dual to the Wigner function. 
This phase-space framework provides two key advantages: analytical tractability which 
simplifies derivation of exact solutions for the full composite system and 
operational efficiency which enables straightforward partial traces over subsystems and 
observable calculations in arbitrary partitions.

The paper is organized as follows: section \ref{model} details the two-qubit-oscillator model and 
describes the dynamics of the qubit-oscillator subsystems. Within this section, we show that 
the fidelity amplitude serves a measure of correlation in a qubit-oscillator system under 
dephasing coupling dynamics, and analyze the redistribution of entanglement in our setup.
In section \ref{probing} we describe a method for probing entanglement 
in the two-oscillator system by indirect measurements performed exclusively 
on the two-qubit system. Finally in section \ref{Summary} we conclude 
with a summary and discussion of results.

\section{\label{model} The two qubit - two oscillator model}
\begin{figure}[!htbp]
  \begin{center}
  \includegraphics[scale=0.5]{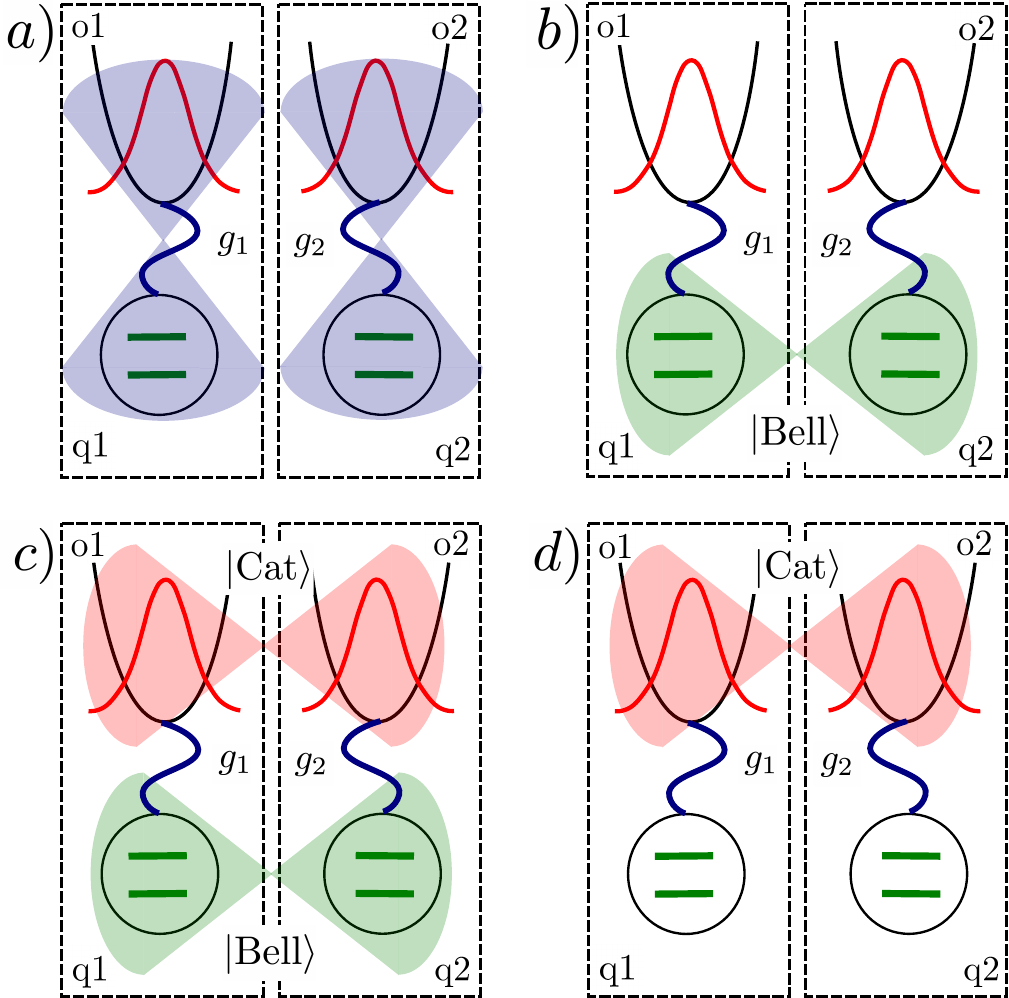}
  \end{center}
  \caption{\label{fig1}  
  Schematic representation of the composite system consisting of two qubits and two harmonic oscillators. 
  Each qubit; represented as a pair of discrete energy levels, interacts with its corresponding oscillator via dephasing coupling. 
  The oscillators are illustrated as Gaussian wave packets confined within quadratic potentials.
  Panel $a)$: initial fully separable configuration of the four-partite system. In this scenario, the dephasing 
  interaction dynamically generates quantum correlations between each qubit and its respective oscillator.  
  Panel $b)$: the two-qubit subsystem is initially prepared in a Bell state, while the two-oscillator subsystem remains separable.  
  Panel $c)$: both the two-qubit subsystem and the two-oscillator subsystem are initially entangled.  
  Panel $d)$: the qubits are initialized in separable coherent superpositions, while the two-oscillator subsystem is entangled.  
  These configurations are central to the indirect entanglement probing protocol explored in this work.}
  \end{figure}

The model comprises two qubit-oscillator subsystems,
where each qubit interacts with its corresponding oscillator via a 
dephasing coupling (Fig. \ref{fig1}).

The Hamiltonian of the system is given by:
\begin{equation}\label{Ham}
    H = \sum_{i=1,2}H_{\text{q}i}+ H_{\text{o}i} + H_{\text{I}i},
\end{equation}
where
 \begin{eqnarray}\nonumber
    H_{\text{q}i} = {\Delta_i\over 2} \sigma_i^{z},\;\,
    H_{\text{o}i} =\omega_i(\h{a}^{\dag}_i\h{a}_i + 1/2),\;\,
    H_{\text{I}i} = g_i\sigma^{z}_{i}\,\h{x}_i,\\\label{Hparts}
\end{eqnarray}
and $\Delta_1 = \omega_{\text{q1}}/\omega_{\text{o1}}$, $\Delta_2 = \omega_{\text{q2}}/
\omega_{\text{o1}}$, $\omega_1 = 1$ and $\omega_2 =\Omega = \omega_{\text{o2}}/\omega_{\text{o1}}$ are 
set in this way in order to place the system in dimensionless units. Additionally, 
$g_i = \lambda_i/\omega_{\text{o1}}$ where $\lambda_i$ represents the interaction strengths 
of the coupling between the qubits and the oscillators. 
By denoting $|g\rangle$ and $|e\rangle$ as the ground or 
excited states in the qubits, the projection of the von-Neumann equation of 
the system into the two-qubit
computational basis states: $|\text{q1},\text{q2}\ra \rightarrow $ 
$|g1g2\ra = |1\ra$, $|g1e2\ra = |2\ra$, $|e1g2\ra = |3\ra$ and
$|e1e2\ra = |4\ra$;  yields a set of decoupled
dynamical equations for the composed two oscillator density operator:
\begin{equation}\label{dmeq}
    \rmi \dot{\varrho}_{ij} = \mathcal{L}_{ij}[\varrho_{ij}],
\end{equation}
where $\mathcal{L}_{ij}[\cdot]$ represents the superoperator acting on
the corresponding density matrix element (see Appendix \ref{app1}).\\

To handle the continuous degrees of freedom of the two-oscillator subsystem, 
we adopt the characteristic function representation~\cite{Ozo98, Ozo02, Oz04}, 
which provides a natural framework for continuous-variable systems. 
The transformation from the density matrix to this phase-space representation 
is given by:

\begin{equation}\label{transfw}
    \rmw_{ij}(\vec{R},t) = \int \rmd \vec{q} \, \varrho_{ij}(\vec{q},\vec{s},t)\,
    \rme^{\rmi \vec{k}\cdot \vec{q}},
    \end{equation}
where $\vec{R} = (k_1,s_1,k_2,s_2)^{T} = (\vec{r}_1,\vec{r}_2)^T$ describes the 
four dimensional position vector in the Fourier phase-space, to which each 
pair $\{k_i,s_i\}$ is associated to a single oscillator, $\vec{q} = (q_1,q_2)^T$, 
$ \vec{k} = (k_1,k_2)^T$, and 
\begin{equation}
\varrho_{ij}(\vec{q},\vec{s},t) = \left\la q_1\!+\!{s_1\over 2}, q_2\!+\!{s_2\over 2}\right| \varrho_{ij}(t)
\left|q_1\!-\!{s_1\over 2}, q_2\!-\!{s_2\over 2}\right\ra .
\end{equation}

This approach allows us to systematically explore all possible dynamical scenarios governed 
by the initial states of the qubits and oscillators. For our purposes, however, 
we focus on initially separable qubit-oscillator pure states:
\begin{eqnarray}\label{ic}
\varrho(\vec{R}_o,t_o) &= &\varrho_{q's}(t_o)\rmw(\vec{R}_o,t_o) \\\nonumber
&&\\\nonumber
&=& \left(\begin{array}{cccc}
c_{11} & c_{12} & c_{13} & c_{14}\\
c_{21} & \hdots\\
\vdots
&&\hdots &c_{44}
\end{array} \right)\,\rmw(\vec{R}_o,t_o)\,,
\end{eqnarray}
with $\vec{R}_o = \vec{R}(t_o)$ and $\rmw(\vec{R}_o,t_o)$ 
being the characteristic function of the two oscillators 
initial conditions and $\tr\varrho^2_{q's}(t_o) = 
\int_{\mathbb{R}^2} \rmd \vec{R}^{\,2} |\rmw(\vec{R}_o,t_o)|^2/(2\pi)^2 = 1 $. 
The dynamics of the full composite system 
to time $t$ is described as (see appendix {\ref{app2}} for the derivation of the 
analytical solution):
\begin{eqnarray}\label{evtot}
    \varrho(\vec{R}, t) &=& \sum_{ij}\rmw_{ij}(\vec{R},t) |i\rangle\langle j| \\\nonumber
    &=& \left(\begin{array}{cccc}
        \rmw_{11}(\vec{R},t) && \hdots & \rmw_{14}(\vec{R},t)\\
        \rmw_{21}(\vec{R},t) & \hdots\\
        \vdots
        &&\hdots &\rmw_{44}(\vec{R},t)
        \end{array} \right),
\end{eqnarray}
while the dynamics of the different subsystems 
is obtained by performing partial traces over the complementary degrees of
freedom. Partial trace over the two-oscillator subsystem is performed
by evaluating the characteristic function variables at the origin,
\ie, the two-qubit reduced system will be obtained by:
  \begin{eqnarray}\label{Nonentqrho}
    \varrho_{\text{q's}}(t) &=&\tr_{\text{o's}} [\varrho(\vec{R},t)]\\\nonumber
    &=& \sum_{ij=1}^4 \rmw_{ij} (\vec{R},t) \big|_{\vec{R}=0} \, |i\rangle\langle j|\,;
  \end{eqnarray}
on the other hand, partial trace over the two-qubit 
degrees of freedom yields the following solution 
for the two-oscillator subsystem: : 
  \begin{eqnarray}\label{oscdyn}
    \rmw_{\text{o's}}(\vec{R},t) &=& \tr_{\text{q's}} [\varrho(\vec{R},t)] \\\nonumber
    &=& \sum_{i=1}^4\rmw_{ii}(\vec{R},t)\,.
  \end{eqnarray}

\subsection{Fidelity amplitude as a quantum correlation probe}
\begin{figure}[!htbp]
  \begin{center}
  \includegraphics[scale=0.38]{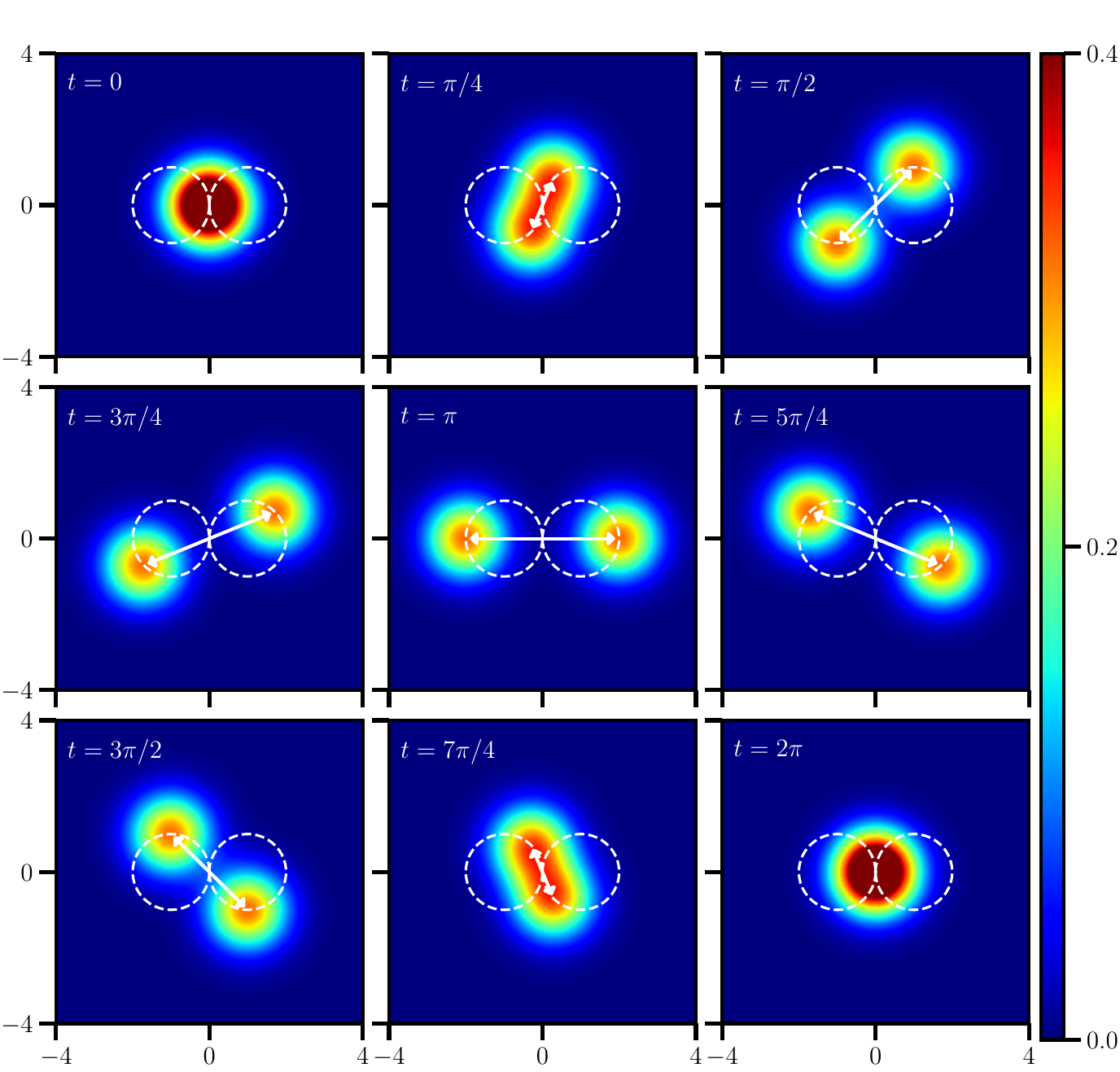}
  \end{center}
  \caption{\label{fig2} Wigner function dynamics showing periodic splitting/recombination of an 
  initial oscillator ground state under qubit coupling ($g=1$). The separation $\vec{d}(t)$ 
  (white arrow) governs both the Gaussian trajectories and fidelity amplitude evolution.
  The panels depicts position (horizontal) versus momentum (vertical) axis.}
\end{figure}

We begin by showing that the fidelity amplitude quantifies qubit-oscillator 
correlations in a single dephasing-coupled subsystem. For doing so let us 
consider a single qubit-oscillator subsystem by tracing out the complementary 
subsystem yielding the following reduced density matrix:
\begin{equation} \label{sqodeph}
  \varrho(\vec{r},t) = \begin{pmatrix}
    \rmw_{ee}(\vec{r},t) & \rmw_{eg}(\vec{r},t)\\
    \rmw_{ge}(\vec{r},t) & \rmw_{gg}(\vec{r},t)
  \end{pmatrix},
\end{equation}
where $\rmw_{ij}(\vec{r},t)$ are derived in Appendix \ref{app1}. 
For a qubit initially in a coherent superposition state: 
$|\psi_{\text{q}}\rangle = 1/\sqrt{2}( |e\rangle +  |g\rangle )$, 
the dephasing coupling dynamics creates periodic 
quantum correlations between the qubit and the oscillator; the Wigner function 
visualization (Fig. \ref{fig2}) of the oscillator reduced dynamics   
reveals this explicitly: an initial ground state splits into two 
counter-propagating Gaussians (separated by $\vec{d}(t)$, depicted as the white 
vector in Fig. \ref{fig2}) that recombine after one period. 

Similarly, the qubit dynamics (from tracing out the oscillator) are:
\begin{equation}\label{rhoq}
  \varrho_{\text{q}}(t) = \frac{1}{2} \begin{pmatrix}
    1 & f_{\text{q}}(t)\\
    f^{*}_{\text{q}}(t) & 1
  \end{pmatrix},
\end{equation}
where $f_{\text{q}}(t) = e^{i\Delta t - |\vec{d}(t)|^2/2}/2$ is the coherence function. 
The fidelity amplitude $|f_{\text{q}}(t)|$~\cite{Go04}, (equivalent to the Loschmidt 
echo~\cite{Go06,Ga97}) quantifies state distinguishability due to coupling from its 
initial configuration.

On the other hand, for bipartite systems of arbitrary dimensions $|\Psi_{\text{AB}}\rangle$,
the I-concurrence connects the degree of mixture of the sub-partitions to the degree of 
entanglement among them:
\begin{equation}\label{Iconc}
  I_{\text{AB}} = \sqrt{ \mathcal{N}\left(1- \tr \varrho^2_{\text{A(B)}} \right)}\,,
\end{equation}
where $\tr \varrho^2_{\text{A(B)}}$ is the purity of the reduced subsystems (A or B), 
and  $\mathcal{N} = {N/ (N -1) }$ (with 
$N=\mathrm{min}(\mathrm{dim}\mathcal{H}_\text{A}, \mathrm{dim}\mathcal{H}_\text{B})$)
ensures $I_{\text{AB}}$ reaches its maximum when the smallest partition is maximally entangled. 
Notably, the measure is invariant under sub-partition choice for pure systems, 
as guaranteed by the Schmidt decomposition~\cite{Nielsen10Ch2}, which equates 
the purities of $\varrho_{\text{A}}$ and $\varrho_{\text{B}}$ regardless of interactions 
among them.

Crucially, the fidelity amplitude relates directly to 
qubit-oscillator entanglement:
\begin{equation}
  I^2_{\text{q}|\text{o}}(t) + |f_{\text{q}}(t)|^2 = 1,
\end{equation}
where $I_{\text{q}|\text{o}}$ is the I-concurrence between the qubit and the oscillator,
see Eq.(\ref{Iconc}), and noticing $\tr\varrho^2_\text{q}=1/2(1-|f_\text{q}(t)|^2)$.
This exact complementarity reveals the fidelity amplitude as a proxy for quantum correlations
in dephasing-coupled systems.

\subsection{\label{entcons} Entanglement redistribution dynamics }

We move forward and consider now the full composite system, where 
the two-qubit subsystem is initially prepared in a Bell 
state,  $|\Psi^+\rangle = 1/\sqrt{2} (|e1g2\rangle +
|g1e2\rangle)$, via the application of a projective operator
$\h{P}=\One_{\text{o's}} \otimes |\Psi^+\rangle\langle\Psi^+|$
to the initial decoupled configuration given in (\ref{ic}).
This is the initial configuration depicted in Fig. \ref{fig1} panel $b)$. 

The system is then allowed to evolve under the dynamics induced by the dephasing model. 
At a later time $t$, the density matrix of the full system becomes:

\begin{eqnarray}\label{solentq}
    \tilde{\varrho}(\vec{R}, t) &=&  {1\over 2} \left(\begin{array}{cccc}
      0 & 0 & 0 & 0\\
      0 & \chi_{+}(\vec{R},t) &  f(\vec{R}, t) & 0\\
      0 & f^{*}(\vec{R}, t)  & \chi_{-}(\vec{R},t) & 0\\
      0 & 0 & 0 & 0
      \end{array} \right)\,, \\
\end{eqnarray}
where the matrix elements $\chi_{\pm}(\vec{R}, t)$ and $f(\vec{R}, t)$
are given by: 
\begin{eqnarray}
  \chi_{\pm}(\vec{R}, t) &=& \rmw\left(\bPh^{-1}(t)\vec{R},t_o\right)
  \, \rme^{\pm \rmi \vec{\delta}(t) \cdot \vec{R} }\,,\\
  f(\vec{R}, t) &=& \rmw\left(\bPh^{-1}(t)\vec{R} + 2\vec{\xi}(t)\,,t_o\right)\,\rme^{\rmi \Delta_{12} \, t} \,,
\end{eqnarray}
with the time-dependent vectors defined as:
\begin{equation}
\vec{\delta}(t) = \int_0^t \rmd t'\, \bPh^T(-t') \vec{\delta},\quad 
\vec{\xi}(t) = \int_0^t \rmd t' \, \bPh(-t') \vec{\xi}, 
\end{equation}
where $\vec{\delta} = (0, g_1, 0, -g_2)^T$, and $\vec{\xi} = (g_1, 0, -g_2, 0)^T$.
These vectors describe the effective displacement of the oscillators in the 
4-dimensional Fourier phase space due to their interaction with the respective qubits.
In the expressions above, $\bPh(t)$ is the transition matrix 
encoding the classical evolution of the two-oscillator system in the dual phase-space 
coordinates and satisfies the group properties: $\bPh(t + s) =\bPh(t)\bPh(s)$,
$\bPh(t=0) =\One$, $\bPh^{-1}(t) =\bPh(-t)$. Explicitly it is written as:
\begin{eqnarray}\label{tmat}
  \bPh(t) &=& \left(
    \begin{array}{cccc}
       \cos(t) & \sin(t) & 0 & 0\\
       -\sin(t) & \cos(t) & 0 & 0\\
       0 & 0 & \cos(\Omega t) & \sin(\Omega t)\\
       0 & 0 & -\sin(\Omega t) & \cos(\Omega t)
     \end{array}\right).
\end{eqnarray}

To track the entanglement in the two-qubit subsystem, we extract 
its reduced density matrix by evaluating the total state at 
$\vec{R}=0$:
\begin{eqnarray}\label{rhobq}
    \tilde{\varrho}_{\text{q's}}(t) &=&  \tilde{\varrho}(\vec{R}, t)\big|_{\vec{R}=0}\\\nonumber
   &=& {1\over 2} \begin{pmatrix}
    0 & 0      & 0    & 0 \\
    0 & 1    & f(t) & 0 \\
    0 & f^*(t) & 1  & 0 \\
    0 & 0      & 0    & 0
    \end{pmatrix},
\end{eqnarray}
with:
\begin{eqnarray}\label{concfun}
f(t) &=&  \rmw\left(2\vec{\xi}(t),t_o\right) \,\rme^{\rmi \Delta_{12} \, t} \,,
\end{eqnarray}
and compute the concurrence $\mathcal{C}$.
For the particular case the state $\varrho_{\text{q's}}$ has $X$-shape form, as
in Eq.~(\ref{rhobq}), according to \cite{Qu12}, its concurrence is simply:
\begin{eqnarray}
    \label{Eq:q1q2-concurrence}
    \mc{C}(t) &=& \abs{f(t)}\,.
\end{eqnarray}

\begin{figure}[!htbp]
  %\hspace*{-0.5cm}
  \begin{center}
  \includegraphics[scale=0.36]{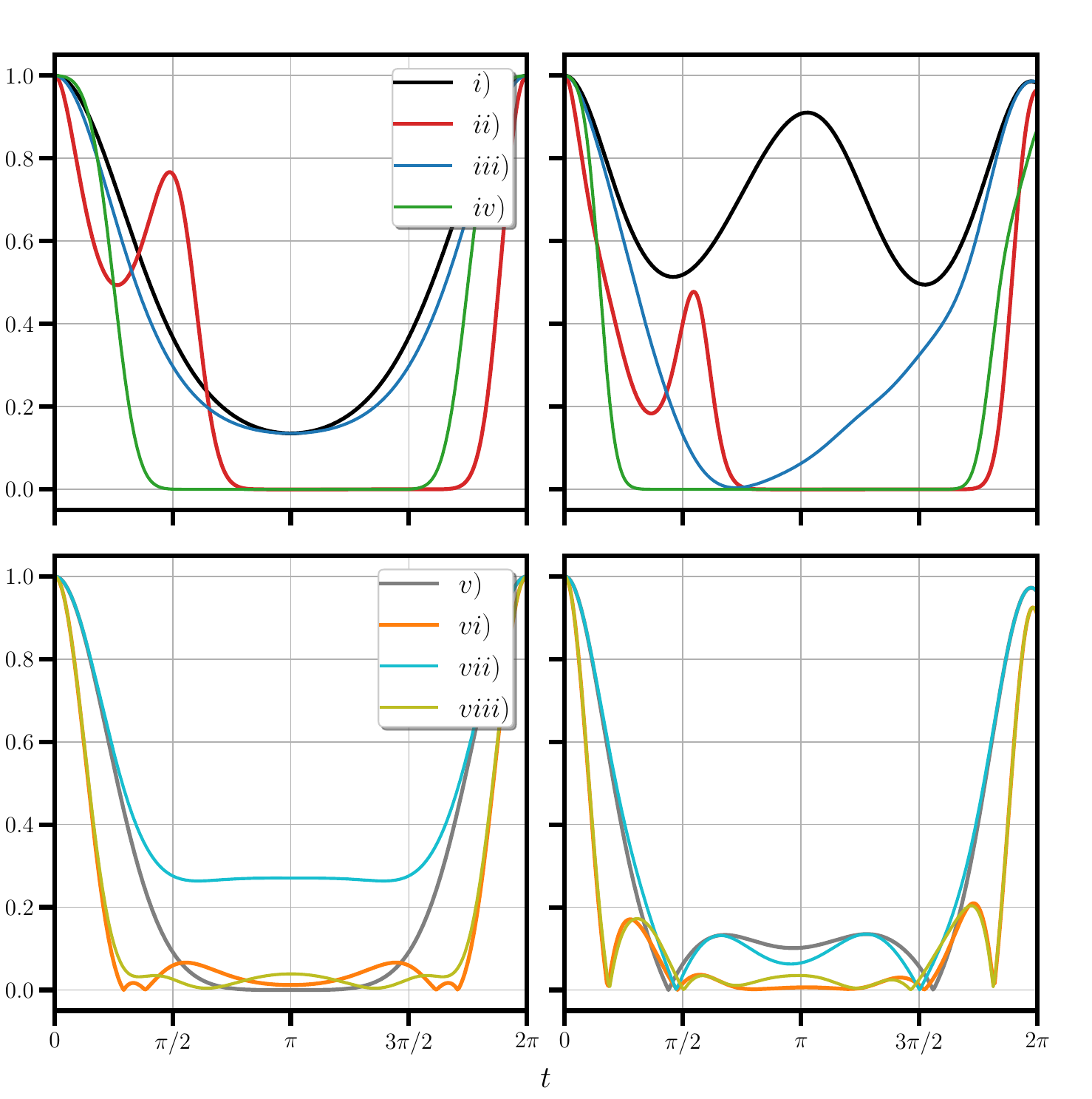}
  \end{center}
  \caption{\label{fig3}  Time evolution of concurrence in the two-qubit subsystem 
for various initial configurations of the two-oscillator subsystem 
(see Appendix~\ref{app2} for detailed descriptions). 
 Top row: 
 i) Separable coherent states with: $\vec{x}_o =(0.5,-0.5,1,-1)^T$ ,  
 ii) Separable single mode squeezed vacuum states with $r=1$, 
 iii) Non-separable cat states with \(\alpha_1 = 1 + i\), \(\beta_1 = -\alpha_1\), \(\alpha_2 = -1 + i\), \(\beta_2 = -\alpha_2\); 
 iv) Two-mode squeeze vacuum state with $r=1$. 
 Bottom row:  
 v) Single excitations separable Fock states: \(|\psi_{\text{o's}}\rangle = (|0\rangle + |1\rangle)(|0\rangle + |1\rangle)/2\),     
 vi) Many excitation separable Fock states:\(|\psi_{\text{o's}}\rangle = (|5\rangle + |2\rangle)(|3\rangle + |1\rangle)/2\),
 vii) Single excitation non-separable Fock states:\(|\psi_{\text{o's}}\rangle = (|10\rangle + |01\rangle)/\sqrt{2}\), 
 viii) Many excitation non-separable Fock states: \(|\psi_{\text{o's}}\rangle = (|51\rangle + |23\rangle)/\sqrt{2}\). 
 At the first column $\omega_1=\omega_1 =1$ and $g_1=g_2=0.5$ while at the second column the asymmetric quasi-periodic regime 
  is depicted: $\omega_1=1$, $\omega_2=\pi$, $g_1=0.5$, $g_2 = \pi/4$. 
}
\end{figure}

Figure~\ref{fig3} illustrates the time evolution of 
$\mathcal{C}(t)$ for various initial states of the two-oscillator subsystem.
In all cases, the concurrence exhibits oscillatory behavior, reflecting the 
exchange of entanglement due to the dephasing interaction. As each qubit 
becomes entangled with its respective oscillator, the bipartite entanglement between 
the qubits fluctuates accordingly. This redistribution of correlations is a direct 
manifestation of the entanglement monogamy principle \cite{CoKuWo00}.

It is possible to quantify the quantum correlations
generated between the two-qubit and the two-oscillator subsystems
using the $I$-concurrence definition from Eq. (\ref{Iconc}) (here 
$\mathcal{N} = 4/3$ ):
\begin{equation}\label{Eq:q1q2-I-concurrence}
    I_{\text{q's}|\text{o's}}(t) = \sqrt{ 2/3  (1 - \abs{f(t)}^2})\,.
\end{equation}
This leads to the following identity, valid for all times $t$:
\begin{equation}\label{Eq:Entanglement-conservation}
      \mc{C}^2(t) + {3 \over 2} \, I_{\text{q's}|\text{o's}}^2(t) = 1\,.
\end{equation}

This result holds independently of the initial state of the 
two-oscillator subsystem. It implies that the rate at which 
the entanglement in the two-qubit subsystem is lost or gained 
is proportional to the rate at which correlations between the qubits and 
the oscillators is respectively gained or lost, reflecting a conserved 
entanglement flux among the parts of the system.

\section{\label{probing}Probing entanglement}

In this section, we propose a method to probe the entanglement properties of 
the two-oscillator subsystem through indirect measurements performed exclusively 
on the two-qubit system. This approach is motivated by two considerations. First, we assume 
that the two-oscillator subsystem is inherently inaccessible to direct measurement, 
making indirect probing techniques essential for characterizing its quantum state. 
Second, our method offers a practical strategy for detecting quantum correlations 
in continuous-variable systems, where standard entanglement measures are often 
challenging to implement due to their mathematical and experimental 
complexity~\cite{Adesso07}.

The proposed method involves preparing two identical copies of the composite system, 
each with the two-oscillator subsystem initialized in the same quantum state. 
In the first copy, the two-qubit subsystem is prepared in the Bell state discussed previously (see panels 
$b)$ and $c)$ in Fig. \ref{fig1}) and the concurrence is tracked during the evolution of the 
dephasing coupling. In the second copy, the two qubits are decoupled and initialized in a coherent 
superposition state (see panels  $a)$ and $d)$ in Fig. \ref{fig1}). This system is also allowed 
to evolve under the same dephasing dynamics. During this evolution, we track the fidelity amplitudes 
of both qubits defined respectively as,
\begin{eqnarray}
  f_1(t) &=& 2| \, \langle e1 | \, \tr_{\text{q2}}\,[\varrho(\vec{R},t)\big|_{\vec{R}=0}] \, |g1\rangle\,| \\\nonumber
         &= &|\rmw\left(2\vec{\nu}(t),t_o\right)|,\\\nonumber
         &&\\
  f_2(t) &=& 2|\, \langle e2|\, \tr_{\text{q1}}\,[\varrho(\vec{R},t)\big|_{\vec{R}=0}]\, |g2\rangle\,| \\\nonumber
         &=&|\rmw\left(2\vec{\mu}(t),t_o\right)| ,
\end{eqnarray}
(see Apendix~\ref{app1} for details of the derivation) as well as the purity of
the reduced two-qubit subsystem.

Crucially, the product of the fidelity amplitudes,
\begin{equation}
\mathcal{F}(t) = f_1(t)f_2(t)\,,
\end{equation} 
which encapsulates the qubit-oscillator correlations generated by the 
dephasing interaction, exactly matches the concurrence $\mathcal{C}(t)$
of the Bell-state-prepared qubit subsystem if and only if 
the two-oscillator subsystem is in a separable (factorizable) state. 
Any deviation from this identity signals the presence of entanglement 
between the oscillators, manifesting as a non-equilibrated redistribution 
of quantum correlations. In summary: 
\begin{eqnarray}\nonumber
\mathcal{C}(t) &=& \mathcal{F}(t), \quad \text{if the oscillator subsystem is separable}, \\\nonumber
\mathcal{C}(t) &\neq& \mathcal{F}(t), \quad \text{otherwise}.
\end{eqnarray}

\begin{figure}[!htbp]
  \begin{center}
  \includegraphics[scale=0.36]{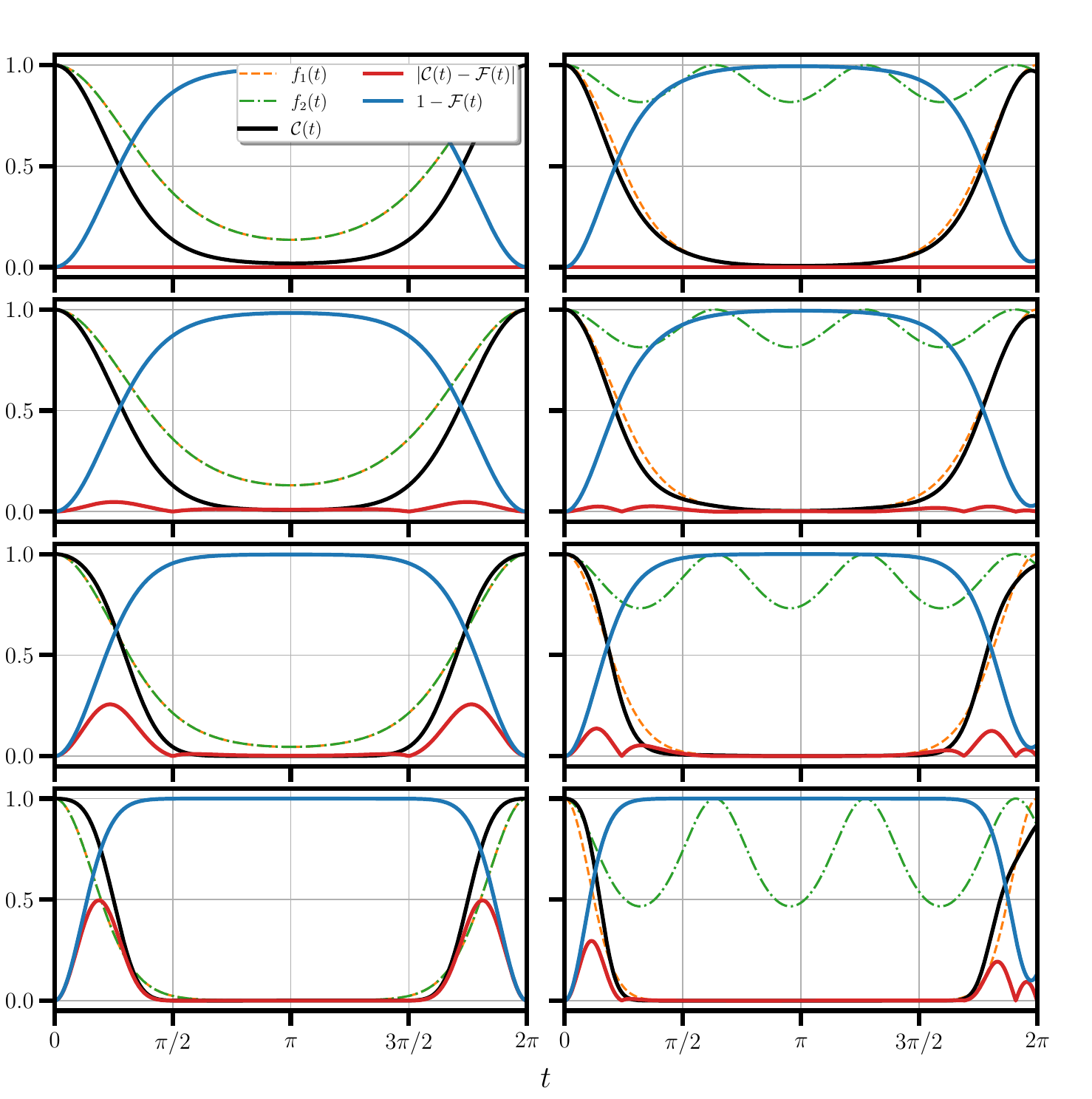}
  \end{center}
  \caption{\label{fig4}  
  Time evolution of the concurrence $\mathcal{C}(t)$ of the two-qubit subsystem 
  initialized in the Bell state, compared with the product of the fidelity 
  amplitudes $\mathcal{F}(t) = f_1(t) f_2(t)$ of two qubits initialized in separable coherent 
  superpositions. The two-oscillator subsystem is initialized in a two-mode squeezed vacuum 
  state (see Appendix~\ref{app2} for details). The squeezing parameter $r$, which controls 
  the amount of entanglement, is varied across rows: $r = 0$ (no entanglement) 
  in the first row, $r = 0.1$ in the second, $r = 0.5$ in the third, and $r = 1$ in the fourth. 
  The left column corresponds to a symmetric regular regime with parameters 
  $\omega_1 = \omega_2 = \Omega = 1$ and $g_1 = g_2 = 0.5$; the right column 
  shows an asymmetric quasi-periodic regime with $\omega_1 = 1$, $\omega_2 = \Omega = \pi$, 
  $g_1 = 0.5\), and \(g_2 = \pi/4$. The blue curve is depicted as such for a better 
  appreciation of the loss of symmetry when quantum correlations among the oscillators 
  are present.}
\end{figure}

\begin{figure}[!htbp]
  \begin{center}
  \includegraphics[scale=0.36]{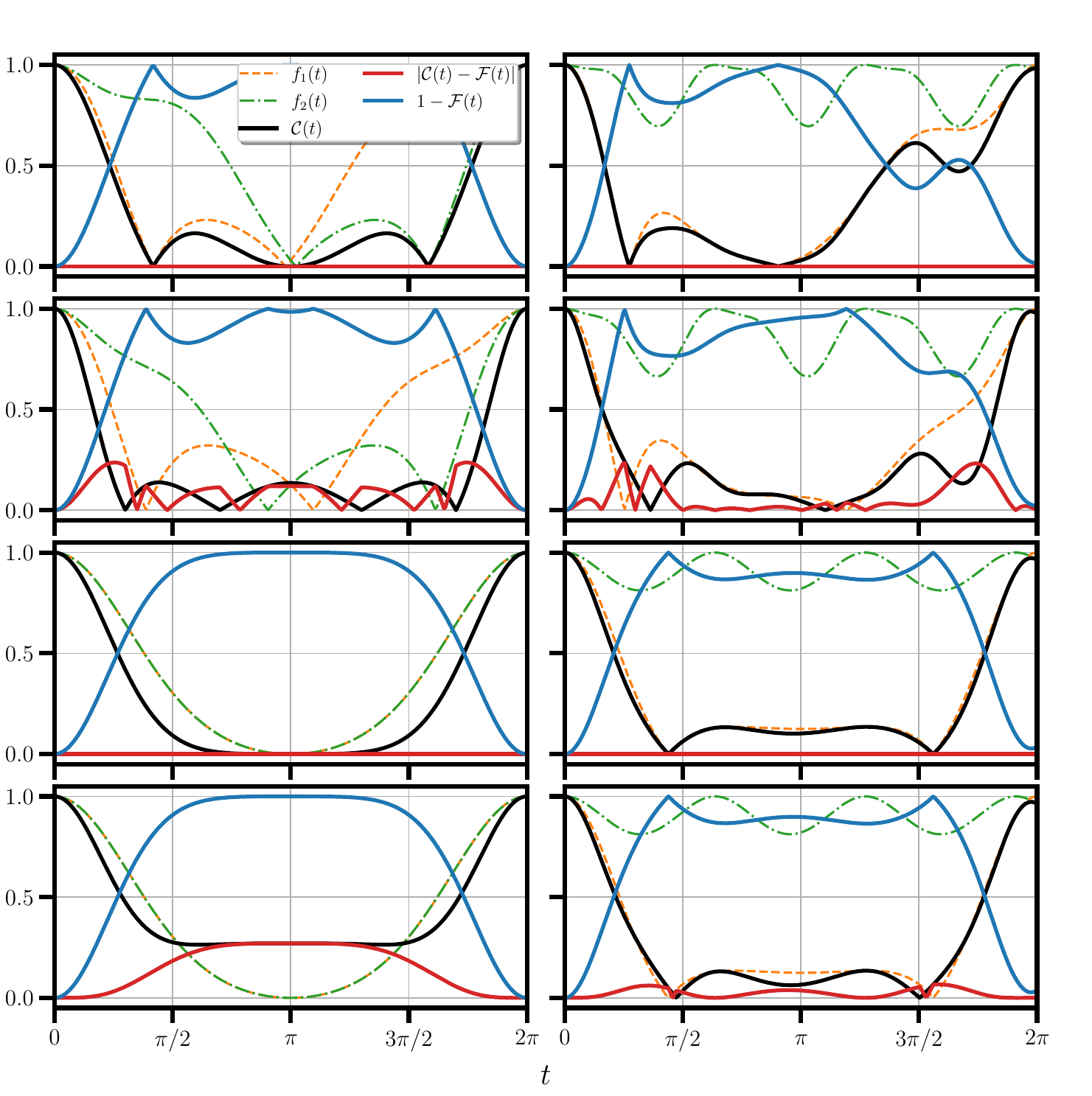}
  \end{center}
  \caption{\label{fig5}  
  Time evolution of the concurrence $\mathcal{C}(t)$ of the two-qubit subsystem initialized 
  in the Bell state, compared with the product of fidelity amplitudes $\mathcal{F}(t) = f_1(t) f_2(t)$ 
  from qubits initialized in separable coherent superpositions. Each row corresponds to a 
  different initial configuration of the two-oscillator subsystem (see Appendix~\ref{app2} for details):  
  first row, separable cat states with $\alpha_1= 1+\rmi$, $\beta_1=-\alpha_1$, 
  $\alpha_2=-1+\rmi$, $\beta_2=-\alpha_2$; second row, entangled cat state
  with the same $\alpha_1$, $\beta_1$, $\alpha_2$, $\beta_2$ as in the separable case;  
  third row, separable superpositions of single-excitation Fock states:
  \(|\psi_{\text{o's}}\rangle = (|0\rangle + |1\rangle)(|0\rangle + |1\rangle)/2\);  
  fourth row, entangled single-excitation Fock state: 
  \(|\psi_{\text{o's}}\rangle = (|10\rangle + |01\rangle)/\sqrt{2}\).
  The parameters employed for this initial configurations are the same used in 
  Fig. \ref{fig3}.  
  The left column uses the symmetric regular configuration 
  $\omega_1 = \omega_2 = \Omega = 1$, $g_1 = g_2 = 0.5$;  
  the right column shows the asymmetric quasi-periodic regime with 
  $\omega_1 = 1$, $\omega_2 = \Omega = \pi$, $g_1 = 0.5$, and $g_2 = \pi/4$.
  The blue curve is depicted as such for a better 
  appreciation of the loss of symmetry when quantum correlations among the oscillators 
  are presen}
\end{figure}

This result is demonstrated in Figures~\ref{fig4} and~\ref{fig5}. 
These figures display the concurrence alongside the individual fidelity amplitudes
$f_1(t),\,f_2(t)$, and the absolute difference $|\mathcal{C}(t) - \mathcal{F}(t)|$
or various initial configurations of the two-oscillator subsystem. As shown, the 
concurrence matches the product of the fidelity amplitudes only when the 
oscillator subsystem is initialized in a separable state.

Figure~\ref{fig4} illustrates this behavior for a two-mode squeezed vacuum state, 
where the degree of entanglement is controlled by the squeezing parameter $r$ 
(see Appendix~\ref{app2}). In Figure~\ref{fig5}, the first two rows depict 
the case of separable and entangled cat states, respectively. The last two rows 
show Fock-state initializations: the third row corresponds to a separable 
superposition of single-excitation Fock states, while the fourth row presents 
an entangled Fock state involving a single excitation.

In this context, qualitative information about the quantum correlations present 
in the inaccessible two-oscillator subsystem can be inferred by comparing the 
dynamics of the concurrence, obtained from the copy where the two-qubit subsystem 
is initially prepared in a Bell state, with the product of the fidelity amplitudes 
recorded from the second copy, in which the qubits evolve independently in 
separable superposition states.

Once the presence of quantum correlations in the two-oscillator subsystem has been confirmed, 
we conjecture that the entanglement flux conserved between the oscillators and the 
qubits; previously discussed in Subsection~\ref{entcons}, is independent of the specific 
sub-partitions and their internal dynamics. We propose that this conservation follows a 
relation analogous to Eq.~(\ref{Eq:Entanglement-conservation}); \ie, the loss or gain 
of quantum correlations within the oscillator subsystem is reflected in the correlations 
established between the two-qubit and two-oscillator subsystems via the dephasing coupling.

To formalize this idea, we introduce an effective quantity $\tilde{\mc{C}}(t)$, 
representing the entanglement content of the two-oscillator subsystem, 
which satisfies the relation:
\begin{equation}\label{Eq:Entanglement-conservation2}
\tilde{\mathcal{C}}^{2}(t) + \mathcal{A}\, I_{\text{q's}|\text{o's}}^2(t) = 1,
\end{equation}
where $\mathcal{A}$ is a proportionality constant. 
Accordingly, by measuring the purity of the two-qubit 
subsystem in the second copy of the system; where the qubits 
are initially decoupled, we can approximate the entanglement 
dynamics of the oscillator subsystem:
\begin{equation}\label{osconc}
\tilde{\mathcal{C}}(t) = \sqrt{\tr[ \varrho^2_{\text{q's}}(t)]}.
\end{equation}

To validate our conjecture, we compare the inferred entanglement measure 
$\tilde{\mc{C}}(t)$ Eq.~(\ref{osconc}) with the logarithmic negativity; 
a widely accepted entanglement measure for continuous-variable Gaussian states. 
Logarithmic negativity is based on the partial transpose of the system's density matrix, 
which, for Gaussian states, translates into a well-defined transformation of 
the covariance matrix. For this comparison, we consider the two-oscillator 
subsystem to be initially prepared in a two-mode squeezed vacuum state:
\begin{equation}
  |\psi_{\text{TMS}}\rangle = \sech(r)\sum_{n=0}^{\infty} \tanh^n(r)|n\rangle_{\text{o1}}|n\rangle_{\text{o2}}\,, 
\end{equation}
where the entanglement between the oscillators is controlled by the squeezing parameter $r$.
The logarithmic negativity $E_N$ for continuous-variable Gaussian states is defined as~\cite{Weedbrook12}:
\begin{equation}
  E_N = \text{max}\left(0,-\log_2\tilde{\nu}_-\right),
\end{equation}
where $\tilde{\nu}_-$ is the smallest symplectic eigenvalue of the partially transposed covariance matrix. 
This is computed from the spectrum of $|\rmi \Omega \sigma^{T_B}|$, 
where $\Omega$ is the symplectic form for two modes: 
$\scriptsize{\Omega = \left(\begin{array}{cc} \omega & 0 \\ 0 & \omega\end{array}\right)}$, with 
$\scriptsize{\omega = \left(\begin{array}{cc} 0 & 1 \\ -1 & 0\end{array}\right)}$. The partially 
transpose covariance matrix $\sigma^{T_B}$ is given by $\sigma^{T_B} = T\, \sigma\,  T$ 
where $\sigma$ is the covariance matrix of the two-oscillator 
subsystem, and $T = \text{diag}(1,1,1,-1)$ implements transposition with respect to the second 
oscillator.

\begin{figure}[!htbp]
  \begin{center}
  \includegraphics[scale=0.36]{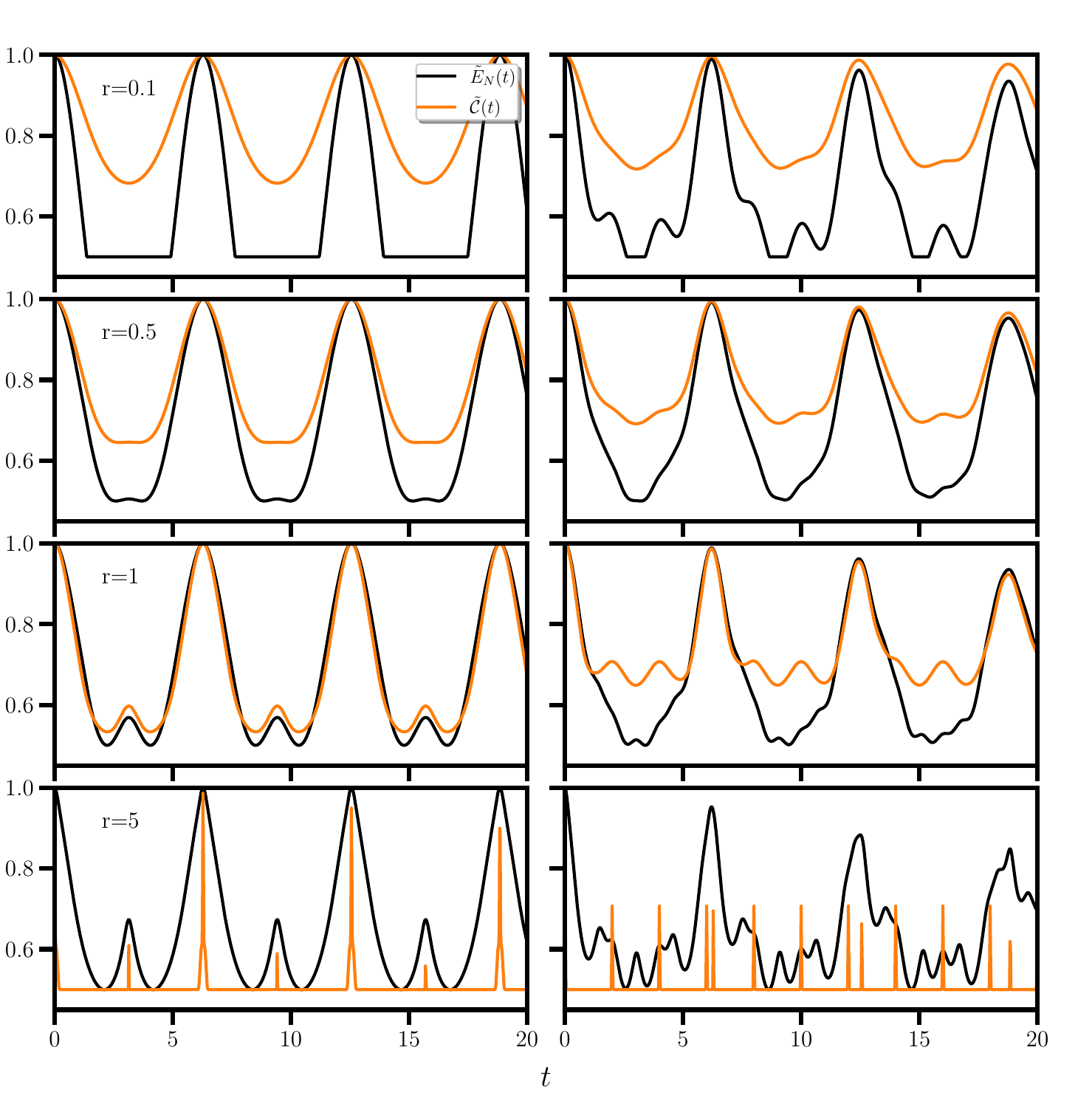}
  \end{center}
  \caption{\label{fig6} 
  Comparison of the dynamical behavior between the logarithmic negativity of 
  the two-oscillator subsystem and the square root of the purity of the two-qubit 
  subsystem, $\tilde{\mathcal{C}}(t) = \sqrt{\tr[\varrho_{\text{q's}}^2(t)]}$. 
  The system is initialized with the two oscillators in a two-mode squeezed vacuum state 
  (see Appendix~\ref{app2} for details), for various values of the entanglement (squeezing) parameter $r$.
  The left column corresponds to a regular regime with parameters $\omega_1 = \omega_2 = \Omega = 1$,
  $g_1 = g_2 = 0.5$; the right column shows an asymmetric quasi-periodic regime with $\omega_1 = 1$,
  $\omega_2 = \Omega = \pi$, $g_1 = 0.5$, and $g_2 = \pi/4$.}
\end{figure}

In Figure~\ref{fig6}, we show the comparison between the normalized logarithmic negativity 
$\tilde{E}_N(t)\in (0.5,1)$ and the inferred measure $\tilde{\mc{C}}(t)$
for different values of the squeezing parameter $r$. As observed, both measures exhibit 
similar qualitative trends during the time evolution. While discrepancies emerge in the 
quasi-periodic regime and at higher entanglement strengths, the overall behavior of 
$\tilde{\mc{C}}(t)$; inferred solely from measurements on the two-qubit subsystem, 
faithfully captures the qualitative dynamics of entanglement in the two-oscillator system.

\begin{figure}[!htbp]
  \begin{center}
  \includegraphics[scale=0.36]{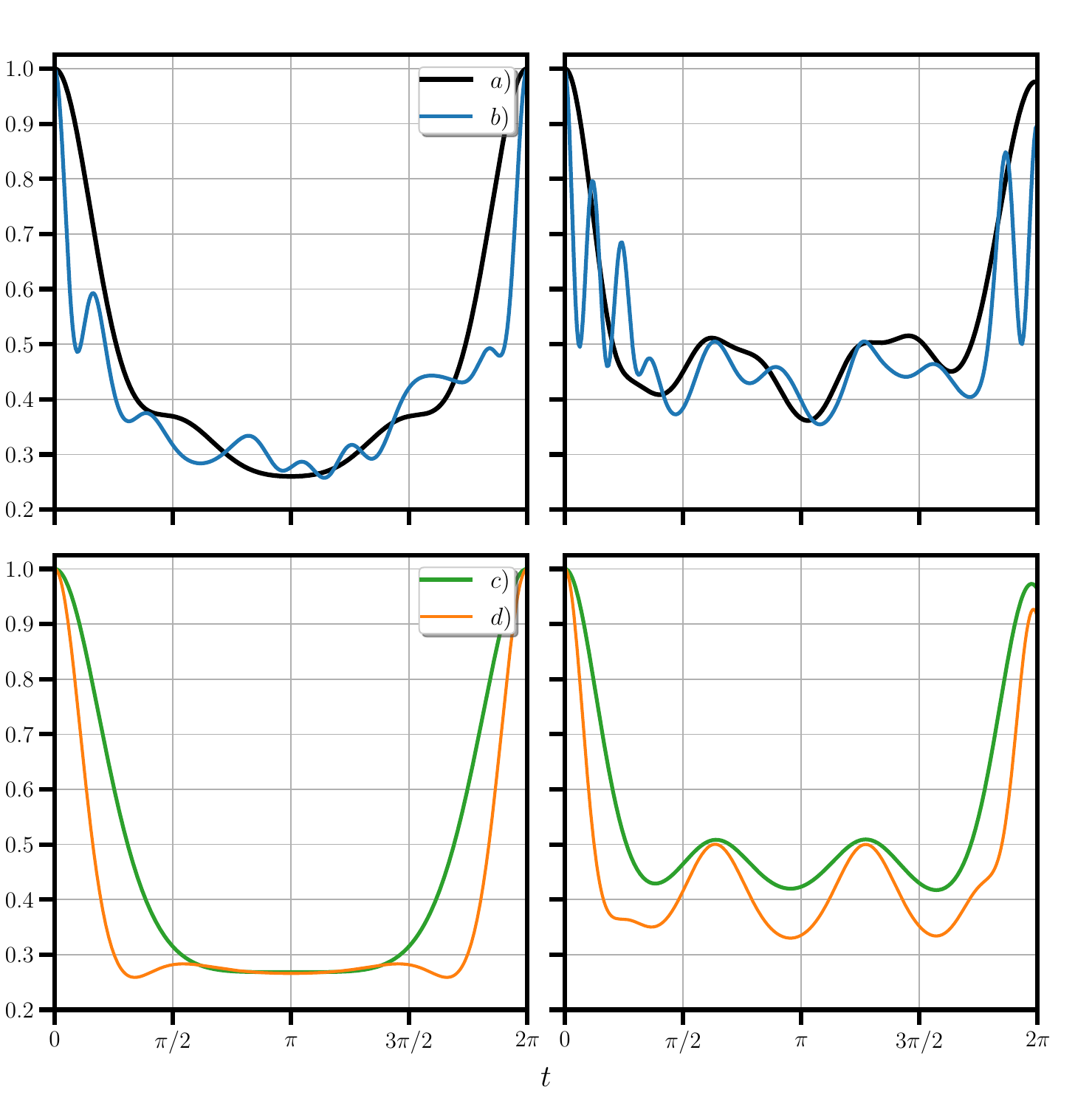}
  \end{center}
  \caption{\label{fig7} 
  Time evolution of \(\tilde{\mathcal{C}}(t) = \sqrt{\tr[\varrho_{\text{q's}}^2(t)]}\) for various entangled initial configurations of the two-oscillator subsystem. 
  label a): entangled cat-state with \(\alpha_1 = 1 + i\), \(\beta_1 = -\alpha_1\), \(\alpha_2 = -1 + i\), \(\beta_2 = -\alpha_2\);  
  label b): entangled cat-state with \(\alpha_1 = 5 + 2i\), \(\beta_1 = -\alpha_1\), \(\alpha_2 = -2 + i/2\), \(\beta_2 = -\alpha_2\);  
  label c): entangled single-excitation Fock state, \(|\psi_{\text{o's}}\rangle = (|10\rangle + |01\rangle)/\sqrt{2}\);  
  label d): entangled many-excitation Fock state, \(|\psi_{\text{o's}}\rangle = (|51\rangle + |23\rangle)/\sqrt{2}\).  
  See Appendix~\ref{app2} for further details on the initial conditions.  
  The top row corresponds to a regular regime with parameters \(\omega_1 = \omega_2 = \Omega = 1\), \(g_1 = g_2 = 0.5\),  
  while the bottom row depicts an asymmetric quasi-periodic regime with \(\omega_1 = 1\), \(\omega_2 = \Omega = \pi\), \(g_1 = 0.5\), and \(g_2 = \pi/4\).}
\end{figure}

Figure~\ref{fig7} presents the time evolution of $\tilde{\mathcal{C}}(t)$ 
for the case in which the two-oscillator subsystem is initially prepared in entangled 
non-Gaussian states. Specifically, we consider both cat-state-like superpositions and entangled Fock states, 
involving single and multiple excitations. These examples demonstrate that $\tilde{\mathcal{C}}(t)$ 
inferred solely from measurements on the two-qubit subsystem, continues to provide qualitative insights 
into the entanglement dynamics of the oscillator subsystem, even beyond the Gaussian regime.

\section{\label{Summary}Summary}
In this work, we have explored the detection of quantum correlations in an 
inaccessible quantum system using a two-qubit quantum probe. Our focus was 
placed on a two-oscillator subsystem initialized in various configurations, 
demonstrating that the proposed probing method is robust and does not depend 
on a specific basis representation.

To infer the quantum correlations within the two-oscillator subsystem, we employed 
a protocol requiring two identical copies of the full system. In the first copy, 
the two-qubit subsystem (the probe) is initialized in a Bell state, and its concurrence 
is tracked throughout the evolution. In the second copy, the qubits are decoupled and 
initialized in coherent superpositions. From this configuration, we measure the fidelity 
amplitudes and the purity of the two-qubit subsystem alone to indirectly retrieve 
information about the entanglement in the inaccessible oscillator subsystem.

A key result of our study is that the concurrence observed in the Bell-state-prepared 
probe exactly matches the product of fidelity amplitudes measured in the second copy; 
if and only if the two-oscillator subsystem is in a separable state. Deviations from 
this correspondence signal the presence of quantum correlations in the oscillator system, 
offering a clear and practical criterion for detecting entanglement through indirect means.

An important and potentially far-reaching observation arising from this study is the 
emergence of an apparent conservation-like behavior of entanglement across the subsystems. 
This redistribution of quantum correlations between the qubit probe and the oscillator 
subsystem throughout the dephasing dynamics suggests the existence of a conserved 
entanglement flux; a property we conjecture may hold more generally for similar 
multipartite systems. While a detailed exploration of this conjecture remains the subject 
of future work, our findings already allow for a qualitative characterization of entanglement 
dynamics in systems where direct measurement is unfeasible.

An essential aspect of our approach is the choice of a two-oscillator subsystem as the 
platform for probing quantum correlations. This system offers a rich variety of configurations 
expressible in a continuous-variable basis and, crucially, is fully integrable; allowing for 
exact analytical treatment of the dynamics and facilitating the identification of clear 
signatures of entanglement. However, it is important to emphasize that the proposed method 
and the diagnostic quantities we use to infer entanglement; such as fidelity amplitudes, concurrence, and purity, 
do not depend on the specific nature of the probed system. This generality suggests that our 
approach may be extended to more complex or generic quantum systems, potentially including 
non-integrable or higher-dimensional setups, thereby broadening its applicability beyond 
the two-oscillator scenario explored in this study.

\acknowledgements
The authors acknowledges financial funding
from Conahcyt (now SECIHTI) through the research project:
Ciencia de Frontera 2019 (No. 10872). LMP also acknowledges 
a doctoral fellowship  from Conahcyt (now SECIHTI).

\appendix

\section{Solutions}\label{app1}
The projection of the von-Neumann equation of the system into the two-qubit computational
basis states as described in (\ref{dmeq}) yields the following set of differential 
equations:
% \begin{eqnarray}\label{rho11}
%   \rmi \dot{\varrho}_{11} &=& [H_{o},\varrho_{11}] - g_1[\h{x}_1,\varrho_{11}]  - g_2[\h{x}_2,\varrho_{11}],\\\nonumber
%   \rmi \dot{\varrho}_{12} &=&\!-\Delta_2 \varrho_{12} \!+\! [H_{o},\varrho_{12}]
%   \!-\! g_1[\h{x}_1,\varrho_{12}]  \!-\! g_2\{\h{x}_2,\varrho_{12}\},\\
%   &&\\\nonumber
%   \rmi \dot{\varrho}_{13} &=&\!-\Delta_1 \varrho_{13}\! + \![H_{o},\varrho_{13}]
%   \!- \!g_1\{\h{x}_1,\varrho_{13}\}  \!-  \!g_2[\h{x}_2,\varrho_{13}],\\
%   &&\\\nonumber
%   \rmi \dot{\varrho}_{14} &=&-(\Delta_1+\Delta_2) \varrho_{14} + [H_{o},\varrho_{14}]\\
%   &&\hspace{1.5cm}- g_1\{\h{x}_1,\varrho_{14}\}  - g_2\{\h{x}_2,\varrho_{14}\},\\\nonumber
%   \\
%   \rmi \dot{\varrho}_{22} &=& [H_{o},\varrho_{22}]
%   - g_1[\h{x}_1,\varrho_{22}]  + g_2[\h{x}_2,\varrho_{22}],\\\nonumber
%   \rmi \dot{\varrho}_{23} &=&-(\Delta_1-\Delta_2) \varrho_{23} + [H_{o},\varrho_{23}]\\
%   &&\hspace{1.5cm}- g_1\{\h{x}_1,\varrho_{23}\}  + g_2\{\h{x}_2,\varrho_{23}\},\\\nonumber
%   &&\\\nonumber
%   \rmi \dot{\varrho}_{24} &=&\!-\Delta_1\varrho_{24} \!+ \![H_{o},\varrho_{24}]
%   \!-\! g_1\{\h{x}_1,\varrho_{24}\}  \!+ \!g_2[\h{x}_2,\varrho_{24}],\\
%   \\
%   \rmi \dot{\varrho}_{33} &=&[H_{o},\varrho_{33}]
%   + g_1[\h{x}_1,\varrho_{33}]  - g_2[\h{x}_2,\varrho_{33}]\\\nonumber
%   \rmi \dot{\varrho}_{34} &=&\!- \Delta_2 \varrho_{34} \!+\! [H_{o},\varrho_{34}]
%   \!+\! g_1[\h{x}_1,\varrho_{34}]  \!-\! g_2\{\h{x}_2,\varrho_{34}\},\\
%   &&\\\label{rho44}
%   \rmi \dot{\varrho}_{44} &=&[H_{o},\varrho_{44}]
%   + g_1[\h{x}_1,\varrho_{44}]  + g_2[\h{x}_2,\varrho_{44}],
% \end{eqnarray}
\begin{equation}\label{rho11}
  \rmi \dot{\varrho}_{11} = [H_{o},\varrho_{11}] - g_1[\h{x}_1,\varrho_{11}]  - g_2[\h{x}_2,\varrho_{11}],
\end{equation}
\begin{equation}
  \rmi \dot{\varrho}_{12} =\!-\Delta_2 \varrho_{12} \!+\! [H_{o},\varrho_{12}]
  \!-\! g_1[\h{x}_1,\varrho_{12}]  \!-\! g_2\{\h{x}_2,\varrho_{12}\},
\end{equation}
\begin{equation}
  \rmi \dot{\varrho}_{13} =\!-\Delta_1 \varrho_{13}\! + \![H_{o},\varrho_{13}]
  \!- \!g_1\{\h{x}_1,\varrho_{13}\}  \!-  \!g_2[\h{x}_2,\varrho_{13}],
\end{equation}
\begin{eqnarray}\nonumber
  \rmi \dot{\varrho}_{14} &=&-(\Delta_1+\Delta_2) \varrho_{14} + [H_{o},\varrho_{14}]\\
  &&\hspace{1.5cm}- g_1\{\h{x}_1,\varrho_{14}\}  - g_2\{\h{x}_2,\varrho_{14}\},
\end{eqnarray}
\begin{equation}
  \rmi \dot{\varrho}_{22} = [H_{o},\varrho_{22}]
  - g_1[\h{x}_1,\varrho_{22}]  + g_2[\h{x}_2,\varrho_{22}],
\end{equation}
\begin{eqnarray}\nonumber
  \rmi \dot{\varrho}_{23} &=&-(\Delta_1-\Delta_2) \varrho_{23} + [H_{o},\varrho_{23}]\\
  &&\hspace{1.5cm}- g_1\{\h{x}_1,\varrho_{23}\}  + g_2\{\h{x}_2,\varrho_{23}\},
\end{eqnarray}
\begin{equation}
  \rmi \dot{\varrho}_{24} =\!-\Delta_1\varrho_{24} \!+ \![H_{o},\varrho_{24}]
  \!-\! g_1\{\h{x}_1,\varrho_{24}\}  \!+ \!g_2[\h{x}_2,\varrho_{24}],
\end{equation}
\begin{equation}
  \rmi \dot{\varrho}_{33} =[H_{o},\varrho_{33}]
  + g_1[\h{x}_1,\varrho_{33}]  - g_2[\h{x}_2,\varrho_{33}],
\end{equation}
\begin{equation}
  \rmi \dot{\varrho}_{34} =\!- \Delta_2 \varrho_{34} \!+\! [H_{o},\varrho_{34}]
  \!+\! g_1[\h{x}_1,\varrho_{34}]  \!-\! g_2\{\h{x}_2,\varrho_{34}\},
\end{equation}
\begin{equation}\label{rho44}
  \rmi \dot{\varrho}_{44} =[H_{o},\varrho_{44}]
  + g_1[\h{x}_1,\varrho_{44}]  + g_2[\h{x}_2,\varrho_{44}],
\end{equation}
where $H_o = H_{o1} + H_{o2}$ are the Hamiltonians of the oscillators as described in (\ref{Hparts}).
Moving into the characteristic function frame can be easily performed by following the following rules
of transformation:
\begin{eqnarray}\label{xpr}
\h{x}^n \h{p}^m\, \varrho &\mapsto&
   \Big( \frac{s}{2} - \rmi \partial_k\Big)^{n}
   \Big( \frac{-k}{2} - \rmi\partial_s\Big)^{m}\! \!\!\!\rmw(k,s),\\
\varrho\, \h{x}^{n}\h{p}^{m} &\mapsto&
   \Big( \frac{-s}{2} - \rmi \partial_k \Big)^n
   \Big( \frac{k}{2} - \rmi\partial_s \Big)^m\! \!\!\! \rmw(k,s),\\
\h{x}^{n}\,\varrho\, \h{p}^{m} &\mapsto&
   \Big( \frac{s}{2} - \rmi \partial_k\Big)^{n}
   \Big( \frac{k}{2} - \rmi\partial_s\Big)^{m}\! \!\!\!\rmw(k,s),\\\label{prx}
\h{p}^{m}\,\varrho\, \h{x}^{n} &\mapsto&
   \Big( \frac{-s}{2} - \rmi \partial_k\Big)^n
   \Big( \frac{-k}{2} - \rmi\partial_s \Big)^m\! \!\!\!\rmw(k,s) ,
\end{eqnarray}
yielding the following set of 1st order partial differential equations:
  \begin{eqnarray}
    \h{L} \rmw_{11}(\vec{R},t) &=& \rmi( g_1 s_1 + g_2 s_2)\rmw_{11}(\vec{R},t),\\
    \h{L}_{12}\rmw_{12}(\vec{R},t) &=& \rmi (\Delta_2 + g_1s_1)\rmw_{12}(\vec{R},t), \\
    \h{L}_{13}\rmw_{13}(\vec{R},t) &=& \rmi (\Delta_1 + g_2s_2)\rmw_{13}(\vec{R},t), \\
    \h{L}_{14}\rmw_{14}(\vec{R},t) &=& \rmi (\Delta_1 + \Delta_2) \rmw_{14}(\vec{R},t), \\\nonumber
    \\
    \h{L} \rmw_{22}(\vec{R},t) &=& \rmi( g_1 s_1 - g_2 s_2)\rmw_{22}(\vec{R},t),\\
    \h{L}_{23}\rmw_{23}(\vec{R},t) &=& \rmi (\Delta_1 - \Delta_2)\rmw_{23}(\vec{R},t), \\
    \h{L}_{24}\rmw_{24}(\vec{R},t) &=& \rmi (\Delta_1 - g_2s_2)\rmw_{24}(\vec{R},t), \\\nonumber
    \\
    \h{L}\rmw_{33}(\vec{R},t) &=& -\rmi( g_1 s_1 - g_2 s_2)\rmw_{33}(\vec{R},t),\\
    \h{L}_{34} \rmw_{34}(\vec{R},t) &=& \rmi( \Delta_2 - g_1 s_1)\rmw_{34}(\vec{R},t),\\\nonumber
    \\
    \h{L} \rmw_{44}(\vec{R},t) &=& -\rmi( g_1 s_1 + g_2 s_2)\rmw_{44}(\vec{R},t),
  \end{eqnarray}
with:
\begin{equation}
    \h{L}  = \partial_t + s_1\partial_{k_1} - k_1\partial_{s_1} + \Omega s_2\partial_{k_2} - \Omega k_2\partial_{s_2},
\end{equation}
while
\[\h{L}_{12} = \h{L} - 2g_2\partial_{k_2},\; \h{L}_{13} = \h{L} - 2g_1\partial_{k_1},\]
\[\h{L}_{14} = \h{L} + 2 g_1\partial_{k_1} - 2 g_2\partial_{k_2},\;
\h{L}_{23} = \h{L} - 2 g_1\partial_{k_1} + 2 g_2\partial_{k_2},\]
\[\h{L}_{24} = \h{L} - 2 g_1\partial_{k_1},\; \h{L}_{34} = \h{L} - 2 g_2\partial_{k_2}.\] 
Now, it is easy to see that all the partial differential equations are
particular cases of a generic differential equation:
\begin{eqnarray}\nonumber
  && \big [ \, \partial_t  +(s_1 + 2\alpha) \partial_{k_1}
  - k_1 \, \partial_{s_1}\\\nonumber
  &&+ (\Omega s_2 + 2\beta ) \partial_{k_2}
  - \Omega k_2 \,\partial_{s_2} \big ] \rmw(\vec{R}, t)\\\nonumber
   && \hspace{3.5cm}= \rmi ( \Delta + \vec{\delta}_{\eps,\zeta}\cdot\vec{R})\, \rmw(\vec{R}, t),
\end{eqnarray}
where $\vec{\delta}_{\eps,\zeta} = (0,\eps,0,\zeta)^T$.
We focus first on solving that generic case and after, we give the specific
values to the involved coefficients regarding the particular cases of the differential 
equations above. The liner partial differential equation can be placed in the 
parametric form:
 \begin{eqnarray}
{\rmd\over\rmd t} \vec{R} (t) &=& {\bf A} \vec{R}(t) + 2\vec{\eta}_{\alpha,\beta},\\
{\rmd \over \rmd t} \rmw(\vec{R}, t) &=& \rmi ( \Delta + \vec{\delta}_{\eps,\zeta} \cdot\vec{R}\,) \rmw(\vec{R}, t),
 \end{eqnarray}
 where 
 \begin{equation}
{\bf A} = \left(
  \begin{array}{cccc}
     0 & 1 & 0 & 0\\
     -1 & 0 & 0 & 0\\
     0 & 0 & 0 & \Omega\\
     0 & 0 & -\Omega & 0
   \end{array}\right)\,,
 \end{equation}
is the stability matrix of the oscillator degrees of freedom and
$\vec{\eta}_{\alpha,\beta} = (\alpha, 0 , \beta, 0)^{T}$. 
The solution for the first of these ordinary differential equations 
is directly obtained:
 \begin{equation}\nonumber
   \vec{R}(t) = \bPh(t-t_o)\vec{R}(t_o) + 2 \vec{\eta}_{\alpha,\beta}(t-t_o),
 \end{equation}
 where
 \begin{equation}\label{xiap}
   \vec{\eta}_{\alpha,\beta}(t-t_o)   =  \int_{t_o}^{t}\rmd t' \bPh(t-t') \vec{\eta}_{\alpha,\beta}.
\end{equation}
This solution has been obtained in terms of the transition
matrix $\bPh$ which is nothing but the exponentiation of the stability
matrix ${\bf A}$:
\begin{equation}
 \bPh(t) = \exp({\bf A} \, t),
\end{equation}
having the following form:
\begin{eqnarray}
  \bPh(t) &=&  \left(
      \begin{array}{cc}
        \bPh_1(t) & 0\\
        0 & \bPh_2(t)
        \end{array}\right)\\\nonumber
  &=&\left(
    \begin{array}{cccc}
       \cos(t) & \sin(t) & 0 & 0\\
       -\sin(t) & \cos(t) & 0 & 0\\
       0 & 0 & \cos(\Omega t) & \sin(\Omega t)\\
       0 & 0 & -\sin(\Omega t) & \cos(\Omega t)
     \end{array}\right)\,,
\end{eqnarray}
and fulfill group properties, \ie:
$\bPh(t + s) =\bPh(t)\bPh(s)$,
$\bPh(t=0) =\One$, $\bPh^{-1}(t) =\bPh(-t)$.

The solution for the second equation can be derived through the employment of the
transition matrix because of its group properties; thus by noticing that
\begin{equation}\label{inmap}
   \vec{R}(t') = \bPh(t'-t)\vec{R}(t) + 2\vec{\eta}_{\alpha,\beta}(t'-t),
\end{equation}
where $\bPh(t'-t) = \bPh^{-1}(t-t')$ and
\begin{eqnarray}
  \vec{\eta}_{\alpha,\beta}(t'-t) &=& - \bPh^{-1}(t-t')\vec{\eta}_{\alpha,\beta}(t-t')\\\nonumber
  &=& - \int_{t'}^t\rmd t'' \bPh(t'-t'')\vec{\eta}_{\alpha,\beta},
\end{eqnarray}
then integration of the second equation can be formulated as:
\begin{equation}
\int_{\rmw(t_o)}^{\rmw(t)}  {\rmd \rmw \over \rmw} = \rmi \int_{t_o}^{t} \rmd t'
 \left\{ \Delta + \vec{\delta}_{\eps,\zeta}\cdot\vec{R}(t')\right\},
\end{equation}
and by employing Eq. (\ref{inmap}):
\begin{eqnarray}
\int_{\rmw(t_o)}^{\rmw(t)}  {\rmd \rmw \over \rmw} &=& \rmi \Delta \cdot (t-t_o)\\\nonumber
&+&\rmi\!\! \int_{t_o}^{t} \!\!\!\!\rmd t'
\vec{\delta}_{\eps,\zeta}\cdot(\bPh(t'-t)\vec{R}(t) +
2\vec{\eta}_{\alpha,\beta}(t'-t))\,.
\end{eqnarray}
Integration of both involved terms can be done using the properties of the
transition matrix; in fact two relevant integrals needed for deriving 
the analytical solutions are integrals of the transition matrix 
$\bPh(t)$ or its inverse $\bPh(-t)$. A straight forward way to calculate these integrals 
is through the differential equations 
which the transition matrix and its inverse satisfies:
\begin{equation}
  {\rmd\over \rmd t} \bPh(t) = \bA \bPh(t),\quad {\rmd\over \rmd t} \bPh^{-1}(t) = -\bPh^{-1}(t) \bA, 
\end{equation}
thus integrating in both sides for the both cases from $t_o$ to $t$; \ie
\begin{eqnarray}
  \int_{t_o}^t \rmd t' {\rmd\over \rmd t'} \bPh(t') &=& \bA \int_{t_o}^t \rmd t'\bPh(t'), \\
  \int_{t_o}^t \rmd t' {\rmd\over \rmd t'} \bPh^{-1}(t') &=& -\int_{t_o}^t \rmd t' \bPh^{-1}(t') \; \bA, 
\end{eqnarray}
yields:
\begin{eqnarray}
  \int_{t_o}^t \rmd t'\bPh(t') &=& \bA^{-1} \left(\bPh(t)- \bPh(t_o) \right), \\
  \int_{t_o}^t \rmd t' \bPh^{-1}(t') &=& - \left(\bPh^{-1}(t) - \bPh^{-1}(t_o) \right)\; \bA^{-1}, 
\end{eqnarray}
and in the case when fixing $t_o=0$ as emplyed in the main text, these becomes:
\begin{eqnarray}
  \int_{0}^t \rmd t'\bPh(t') &=& \bA^{-1} \left(\bPh(t)- \One \right) \\
  \int_{0}^t \rmd t' \bPh^{-1}(t') &=& - \left(\bPh^{-1}(t) - \One \right)\; \bA^{-1}. 
\end{eqnarray}

The solution of the first order partial differential equation is therefore given as:
\begin{eqnarray}\nonumber
  \rmw (\vec{R},t) &=& \rmw (\bPh^{-1}(t-t_o)\vec{R} + 2\vec{\eta}_{\alpha,\beta}(t_o-t),t_o)
  \,\rme^{\rmi \Delta\, (t-t_o)} \\
  && \exp\left(\rmi \vec{\delta}_{\eps,\zeta}(t-t_o) \cdot \vec{R} + \rmi \Gamma^{\eps,\zeta}_{\alpha,\beta}(t-t_o)\right)   
\end{eqnarray}
where 
\begin{equation}
  \vec{\delta}_{\eps,\zeta}(t-t_o) = \int_{t_o}^{t}\rmd t' \bPh^{T}(t'-t) \vec{\delta}_{\eps,\zeta},
\end{equation}
\begin{equation}
  \Gamma^{\eps,\zeta}_{\alpha,\beta}(t-t_o) = 2\int_{t_o}^{t} \rmd t'\, \int_{t_o}^{t'} \rmd t''\, 
  \bPh^{T}(t'-t'')\, 
  \vec{\delta}_{\eps,\zeta} \cdot \vec{\eta}_{\alpha,\beta},
\end{equation}
and we have made the substitution of the 
initial condition corrdinates $\vec{R}(t_o) \rightarrow \vec{R} = \vec{R}(t)$, 
through the map given at (\ref{inmap}); in other words: 
\[ \rmw (\vec{R}(t_o),t_o) \rightarrow \rmw (\bPh^{-1}(t-t_o)\vec{R} + 2\vec{\eta}_{\alpha,\beta}(t_o-t),t_o).\] 
With this general solution we can now writte the solutions 
for the different elements of the system of differential equations and hence the 
time evolution of the elements of the composed system described in (\ref{evtot}): 
  \begin{eqnarray}\nonumber
    \rmw_{11} (\vec{R},t) & = & c_{11}\rmw \left( \bPh^{-1}(t-t_o)\vec{R}\,, t_o\right)
    \rme^{\rmi  \vec{\delta}_+(t-t_o)\cdot \vec{R}}\\\label{w11}
  \end{eqnarray}
  \begin{eqnarray}\nonumber
    \rmw_{12} (\vec{R},t) &=& c_{12}\rmw\left(\bPh^{-1}(t-t_o)\vec{R} + 2\vec{\mu}(t-t_o)\,,t_o\right)\,\\
    && \rme^{\rmi \Delta_2\, (t-t_o) + \rmi \vec{d}_1(t-t_o)\cdot \vec{R}  - \rmi \theta (t-t_o) }
  \end{eqnarray}
    \begin{eqnarray}\nonumber
    \rmw_{13} (\vec{R},t)  &=& c_{13}\rmw\left(\,\bPh^{-1}(t-t_o)\vec{R} + 2\vec{\nu}(t-t_o) \,,t_o \right)\, \\
    &&\rme^{\rmi \Delta_1\, (t-t_o) + \rmi\vec{d}_2(t-t_o)\cdot\vec{R}
    - \rmi \varphi(t-t_o)}
    \end{eqnarray}
    \begin{eqnarray}\nonumber
    \rmw_{14} (\vec{R},t) &=& c_{14}\rmw\left(\bPh^{-1}(t-t_o)\vec{R} - 2 \vec{\xi}(t-t_o) \,,t_o \right)\,\\
    &&\rme^{\rmi \Delta_{12}\, (t-t_o)}
    \end{eqnarray}
    \begin{eqnarray}\nonumber
    \rmw_{22}(\vec{R},t) &=& c_{22}\rmw\left(\bPh^{-1}(t-t_o)\vec{R}\,,t_o\right)\,\rme^{\rmi \vec{\delta}_-(t-t_o)\cdot \vec{R}}\\
    \end{eqnarray}
    \begin{eqnarray}\nonumber
    \rmw_{23} (\vec{R},t) &=& c_{23}\rmw\left(\bPh^{-1}(t-t_o)\vec{R} + 2\vec{\xi}(t-t_o)\,,t_o\right)\\
    &&\rme^{\rmi \Delta_{12}\, (t-t_o)}
    \end{eqnarray}
    \begin{eqnarray}\nonumber
    \rmw_{24} (\vec{R},t) &=& c_{24}\rmw\left(\bPh^{-1}(t-t_o)\vec{R} + 2\vec{\nu}(t-t_o)\,,t_o\right)\,\\
    &&\rme^{\rmi \Delta_1\, (t-t_o) - \rmi \vec{d}_2(t-t_o)\cdot \vec{R} + \rmi \varphi(t-t_o)}
    \end{eqnarray}
    \begin{eqnarray}\nonumber
    \rmw_{33}(\vec{R},t) &=& c_{33}\rmw\left(\bPh^{-1}(t-t_o)\vec{R}\,,t_o\right)\,
   \rme^{- \rmi \vec{\delta}_{-}(t-t_o) \cdot \vec{R} }\\
   \end{eqnarray}
   \begin{eqnarray}\nonumber
    \rmw_{34} (\vec{R},t) &=& c_{34}\rmw\left(\bPh^{-1}(t-t_o)\vec{R}
    + 2\vec{\mu}(t-t_o)\,,t_o\right)\,\\
    &&\rme^{\rmi \Delta_2\, (t-t_o) - \rmi \vec{d}_1(t-t_o)\cdot \vec{R}
    + \rmi \theta(t-t_o) }
    \end{eqnarray}
    \begin{eqnarray}\nonumber
    \rmw_{44}(\vec{R},t)  &=& c_{44}\rmw\left( \bPh^{-1}(t-t_o)\vec{R}\,,t_o\right)\,
   \rme^{- \rmi  \vec{\delta}_+(t-t_o) \cdot \vec{R}}\\\label{w44}
  \end{eqnarray}
where for simplicity we've defined the following functions:
\begin{eqnarray}
  \vec{\mu}(t-t_o) &=& \vec{\eta}_{0,-g_2}(t_o-t)\\\nonumber
                   &=& \int_{t_o}^{t}\rmd t' \bPh^{-1} (t'-t_o) \vec{\eta}_{0,g_2}\\
  \vec{\nu}(t-t_o) &=& \vec{\eta}_{-g_1,0}(t_o-t)\\\nonumber
                   &=&\int_{t_o}^{t}\rmd t' \bPh^{-1} (t'-t_o) \vec{\eta}_{g_1,0}\\
  \vec{\xi}(t-t_o) &=& \vec{\eta}_{-g_1,g_2}(t_o-t)\\\nonumber
                   &=&\int_{t_o}^{t}\rmd t' \bPh^{-1} (t'-t_o) \vec{\eta}_{g_1,-g_2}
\end{eqnarray}
together with:
\begin{eqnarray}
  \vec{\delta}_+(t) = \vec{\delta}_{g_1,g_2}(t), &\;& \vec{\delta}_-(t) = \vec{\delta}_{g_1,-g_2}(t),\\
  \vec{d}_1(t) = \vec{\delta}_{g_1,0}(t),&\;& \vec{d}_2(t) = \vec{\delta}_{0,-g_2}(t),\\
  \theta (t) = \Gamma^{g_1,0}_{0,g2} (t), &\;& \varphi(t) = \Gamma^{0,g_2}_{g_1,0}(t)\,.
\end{eqnarray}

\section{\label{app2}
Oscillators initial conditions in the characteristic function
representation}
Along the main part of the paper several references are given 
about the two-oscillator initial condition configuration. In this 
appendix we describe them and give their explicit form 
in the characteristic function description.\\ 

{\bf Gaussian-separable state}:
In the wave function description these states 
are described as:
\begin{eqnarray}
  \psi(x,y) &=& \psi_1(x)\,\psi_2(y) \\\nonumber
&=&{1\over \sqrt{\pi}}\rme^{\rmi p_{o1} x - (x-x_{o1})^2/
2\sigma^2_{\text{o1}} }\rme^{\rmi p_{o2} y - (y-x_{o2})^2/2\sigma^2_{\text{o1}}}.
\end{eqnarray}
where $x_{\text{o}i}$, and $p_{\text{o}i}$ for $i=1,2$ describes the
initial position and momentum of each oscillator, while $\sigma_{\text{o}i}$ 
is the corresponding width of the wave functions.
This state transform to the characteristic function as:
\begin{eqnarray}
  \rmw(\vec{R},t_o) &=&\rmw(\vec{r}_1,t_o) \rmw(\vec{r}_2,t_o)\\\nonumber
  &=& \exp\left( \rmi \vec{R}\cdot \vec{x}_o - {1\over 2} \vec{R}^T \sigma \vec{R}\right)
\end{eqnarray}
where $\rmw(\vec{r}_i,t_o) = \rme^{\rmi \vec{r}_i\cdot \vec{x}_i-{1\over 2} \vec{r}^T_{i} \sigma_i \vec{r}_i}$
and $\vec{x}_{i} = (x_{\text{o}i},p_{\text{o}i})^T$, $\vec{x}_{\text{o}} = (x_{\text{o1}},p_{\text{o1}},x_{\text{o2}},p_{\text{o2}})^T$,  
while $\sigma_i$ are the correspondent covariance matrices while $\sigma$ is the two-oscillator 
covariance matrix, (the case when $\sigma_i = 1/2\One$
refers to coherent states).\\

{\bf Coherent separable and entangled cat-state:}
A coherent separable cat-state refers to a two 
factorizable cat-state like of each oscillator: 
\begin{equation}
    |\psi_{\text{o's}}(t_o)\ra = (a_1|\alpha_1\rangle +b_1|\beta_1\rangle)
    ( a_2|\alpha_2\ra + b_2 |\beta_2\ra).
\end{equation}
where $a^2_i + b_i^2 = 1$, and each $|\alpha_i\ra$ or $|\beta_i\ra$ represents a coherent state with the 
folowing wave function representation:
\begin{equation}
  \psi_{\alpha}(x) = \la x |\alpha\ra = \rme^{\rmi p_{\alpha} x - (x-x_{\alpha})^2/2}/\pi^{1/4}.
\end{equation}
where $x_{\alpha} = (\alpha + \alpha^*)/2 $ and $p_{\alpha} = (\alpha - \alpha^*)/2\rmi$ 
represents the initial position and momentum of the coherent wave packet.
Each of the oscillators cat-state is described in the characteristic 
function representation as:
 \begin{eqnarray}
    \rmw(\vec{r}_i,t_o) =  \rme^{-  {r_i^2/ 4} }\,
    \bigg(c_i\rme^{\rmi \vec{x}^{\,(i)}_1\cdot\vec{r}_i} &+&
     d_i\rme^{\rmi \vec{x}^{\,(i)}_2\cdot\vec{r}_i}\\\nonumber
      &+& \gamma_i \rme^{{i\over 2} \vec{\eta}_i \cdot \vec{r}_i } +
     \gamma^*_i \rme^{{i\over 2} \vec{\eta}^{\,*}_i\! \cdot \vec{r}_i}\bigg)
  \end{eqnarray} 
where $c_i = |a_i|^2\mathcal{N}$, $d_i=|b_i|^2\mathcal{N}$, 
$\gamma_i = a_ib^*_i\mathcal{N}\rme^{-\zeta_i/4}$ with $\mathcal{N}$ being a normalization 
constant (\ie $\mathcal{N} = \rmw(\vec{r}_i=0,t_o)$), $\vec{x}^{\,(i)}_1=(x_{\alpha_i},p_{\alpha_i})^T$, 
$\vec{x}^{\,(i)}_2=(x_{\beta_i},p_{\beta_i})^T$ and 
  \begin{eqnarray}
  \vec{\eta}_i &=& \left(\begin{array}{c}
                  x_{\alpha_1} + x_{\beta_1} + \rmi(p_{\alpha_1}-p_{\beta_1})\\ 
                  p_{\alpha_1} + p_{\beta_1} - \rmi(x_{\alpha_1}-x_{\beta_1})
                \end{array}\right),\\\nonumber
  \end{eqnarray}
and $\zeta_i = (x_{\alpha_1}\!-\!x_{\beta_1})^2 + (p_{\alpha_1}\!-\!p_{\beta_1})^2  
-2\rmi \,(x_{\alpha_1} + x_{\beta_1})(p_{\alpha_1}-p_{\beta_1})$.
The two-oscillator system each in a cat-state separable configuration is therefore 
given by:
\begin{equation}
  \rmw(\vec{R},t_o) =\rmw(\vec{r}_1,t_o) \rmw(\vec{r}_2,t_o),
\end{equation}
with each $\rmw(\vec{r}_i,t_o)$ describing the individual cat-state oscillator.
A coherent entangled cat-state of the two-oscillator subsystem is defined as:
\begin{equation}
    |\psi_{\text{o's}}(t_o)\ra = c_1|\alpha_1,\beta_2\ra + c_2 |\beta_1,\alpha_2\ra\,.
\end{equation}
 The corresponding form in the characteristic function description is given by:
  \begin{eqnarray}\nonumber
    \rmw(\vec{R},t_o) &=&  \rme^{-  {R^2\over 4} }\,\bigg(a\rme^{\rmi \vec{x}_1\!\cdot\!\vec{R}} \!+\!
     b\rme^{\rmi \vec{x}_2\cdot\!\vec{R}}\! +\! \gamma \rme^{{i\over 2} \vec{\eta} \cdot \vec{R} } \!+\!
     \gamma^* \rme^{{i\over 2} \vec{\eta}^{\,*}\! \cdot \vec{R}}\bigg)\\
  \end{eqnarray}
  where $a = |c_1|^2\mathcal{N}$, $b=|c_2|^2\mathcal{N}$, 
  $\gamma = c_1c^*_2\mathcal{N}\rme^{-\zeta/4}$
  with $\mathcal{N}$ being a normalization 
  constant (\ie $\mathcal{N} = \rmw(\vec{R}=0,t_o)$)
  and  
  \begin{eqnarray}
  \vec{x}_1 &=& \left(\begin{array}{c}
                 x_{\alpha_1}\\p_{\alpha_1} \\ x_{\beta_2}\\ p_{\beta_2}
  \end{array}\right),\;
  \vec{x}_2 = \left(\begin{array}{c}
                x_{\beta_1} \\ p_{\beta_1} \\ x_{\alpha_2} \\ p_{\alpha_2}
  \end{array}\right),\\\nonumber
  &&\\
  \vec{\eta} &=& \left(\begin{array}{c}
                  x_{\alpha_1} + x_{\beta_1} + \rmi(p_{\alpha_1}-p_{\beta_1})\\ 
                  p_{\alpha_1} + p_{\beta_1} - \rmi(x_{\alpha_1}-x_{\beta_1})\\
                  x_{\beta_2} + x_{\alpha_2} + \rmi(p_{\beta_2}-p_{\alpha_2})\\
                  p_{\beta_2} + p_{\alpha_2} - \rmi(x_{\beta_2}-x_{\alpha_2})
                \end{array}\right),\\\nonumber
  \end{eqnarray}
\begin{eqnarray}\nonumber
    \zeta \!\!\!&=&\!\!\! (x_{\alpha_1}\!-\!x_{\beta_1})^2 \!+\! (x_{\beta_2}\!-\!x_{\alpha_2})^2 \!+ \!
            (p_{\alpha_1}\!-\!p_{\beta_1})^2 \!+ \!(p_{\beta_2}\!-\! p_{\alpha_2})^2 \\\nonumber
        &-&\!\!\!2\rmi \left(\,(x_{\alpha_1} + x_{\beta_1})(p_{\alpha_1}-p_{\beta_1}) + 
        (x_{\beta_2} + x_{\alpha_2})(p_{\beta_2} - p_{\alpha_2})\, \right)\,.
\end{eqnarray}

{\bf Separable and entangled Fock states:}
Let us first consider a separable superposition of number states in each oscillator:
\begin{eqnarray}
  |\psi_{osc's}\ra &=& |\psi_{o1}\ra|\psi_{o2}\ra\\\nonumber
  &=& \left(a_1 |n_1\ra + b_1 | m_1\ra \right)\,\left(a_2 |n_2\ra + b_2 | m_2\ra \right)
\end{eqnarray}
with $n_i\neq m_i$ being arbitrary number states, while
$a^2_i + b_i^2 = 1$. In the characteristic function description,
these states become:
\begin{eqnarray}
  \rmw(\vec{R},t_o) &=& \rmw(\vec{r}_1,t_o )\, \rmw(\vec{r}_2,t_o) \\\nonumber
  &=& \rme^{-r^2_1/4}\left(\alpha_1 L_{n1}(r_1^2/2) + \beta_1 L_{m1}(r_1^2/
  2)   \right)\\\nonumber
  &&\rme^{-r^2_2/4}\left(\alpha_2 L_{n2}(r_2^2/2) + \beta_2 L_{m2}(r_2^2/
  2)   \right)\,.
\end{eqnarray}

Now we consider initial entangled number-states
of the oscillators, described by a vector state in the form:
\begin{equation}
|\psi_{osc's}(t_o)\ra = p_{1}|n_1,m_2\ra + p_{2}|m_1,n_2\ra
\end{equation}
which has the following representation in the characteristic function description
(the following expression is obtained for 
the conditions: $n_1-m_1>-1$ and $n_2-m_2>-1$):
\begin{widetext}
\begin{eqnarray}
\rmw(\vec{R},t_o) &=& \rme^{-R^2/4}\bigg\{
|p_1|^2 L_{n_1}(r_1^2/2)L_{m_2}(r_2^2/2) + |p_2|^2 L_{m_1}(r_1^2/2)L_{n_2}(r_2^2/2)\\\nonumber
&& +B\, \bigg [ p_1p_2^* \left( \rmi k_1 + s_1 \over 2\right)^{n_1-m_1}
\left(\rmi k_2 - s_2 \over 2 \right)^{n_2-m_2} L_{m_1}^{(n_1-m_1)}(r_1^2/2)L_{m_2}^{(n_2-m_2)}(r_2^2/2)\\\nonumber
&&\hspace{0.6cm}+ p_1^*p_2 \left(\rmi k_1 - s_1 \over2 \right)^{n_1-m_1}\left(\rmi k_2+s_2\over 2\right)^{n_2-m_2} L_{m_1}^{(n_1-m_1)}(r_1^2/2)L_{m_2}^{(n_2-m_2)}(r_2^2/2)\bigg]
  \bigg\}
\end{eqnarray}
\end{widetext}

{\bf Separable single-squeeze vacuum states and two-mode 
squeeze vacuum state:} The state vector of a single-squeeze vacuum state 
is defined as (with $\theta = \pi/2$):  
\begin{equation}
  |\psi_{\text{SMS}}\rangle = {1\over \sqrt{\cosh(r)}}\sum_{n=0}^{\infty} \sqrt{{(2n)!\over 2^n n!}}(-\rmi \tanh(r) )^n |2n\rangle.
\end{equation}
In the characteristic function description, squeeze states are Gaussian states with mean zero and an uneven  
width in position and momentum, characterized through its covariance matrix. For this particular case, 
the characteristic function description of the two-oscillator subsystem is written as: 
\begin{eqnarray}
  \rmw(\vec{R},t_o) &=& \rmw(\vec{r}_1,t_o)\rmw(\vec{r}_2,t_o)\\\nonumber
  &=& \rme^{-{1\over 2}\vec{r}_1^{\,T}\,\sigma_1 \,\vec{r}_1 } \rme^{-{1\over 2}\vec{r}_2^{\,T}\,\sigma_2 \,\vec{r}_2 }
\end{eqnarray} 
where for simplcity we consider two identical covariance matrices:
\begin{equation}
  \sigma_1 = \sigma_2 = {1\over 2}\left(\begin{array}{cc}
    \cosh(2r) & -\sinh(2r)\\
   -\sinh(2r) & \cosh(2r) 
  \end{array}\right)
\end{equation}  

A two-mode sqeeze vacumm state is defined as: 
\begin{equation}
  |\psi_{\text{TMS}}\rangle = \sech(r)\sum_{n=0}^{\infty} \tanh^n(r)|n\rangle_{\text{o1}}|n\rangle_{\text{o2}}
\end{equation}
and its representation in the characteristic function description is again Gaussian:
\begin{equation}
  \rmw(\vec{R},t_o) = \rme^{-{1\over 2}\vec{R}^{\,T}\,\sigma \,\vec{R} }
\end{equation} 
although now, entanglement is characterized through correlations appearing in the composite system 
covariance matrix: 
\begin{equation}
  \sigma = {1\over 2}\left(\begin{array}{cccc}
    \cosh(2r) & 0 & \sinh(2r)& 0\\
     0 & \cosh(2r) & 0 &-\sinh(2r)\\
     \sinh(2r) & 0 & \cosh(2r) & 0\\
     0 &  -\sinh(2r) & 0 & \cosh(2r)
  \end{array}\right)\,.
\end{equation}

\bibliography{man}
\end{document}